\documentclass[
reprint,
superscriptaddress,
prl,
amsmath,amssymb,
aps,
footinbib
]{revtex4-2}

\usepackage{graphicx}
\usepackage{xr-hyper}
\usepackage{hyperref}
\usepackage{xcolor}
\hypersetup{colorlinks=true, linkcolor=blue, citecolor=blue, urlcolor=blue}
\usepackage{lipsum}
\usepackage[caption=false]{subfig}
\usepackage{mathtools}
\usepackage[normalem]{ulem}

\newcommand{\ka}{{\overline{K}}}
\newcommand{\ks}{{\sigma_K}}
\newcommand{\kij}{{K_{ij}}}

\newcommand{\rbf}{\mathbf{r}}
\newcommand{\pbf}{\hat{\mathbf{p}}}

\newcommand{\RI}{\sigma_I}

\newcommand\underrel[3][]{\mathrel{\mathop{#3}\limits_{%
			\ifx c#1\relax\mathclap{#2}\else#2\fi}}}

\begin{document}

\title{Disordered Yet Directed: The Emergence of Polar Flocks with Disordered Interactions}
\date{\today}

\author{Eloise Lardet}
\affiliation{Department of Mathematics, Imperial College London, 180 Queen's Gate, London SW7 2AZ, United Kingdom}
\author{Rapha\"el Voituriez}
\affiliation{Laboratoire Jean Perrin, UMR 8237 CNRS, Sorbonne Universit\'{e}, 75005 Paris, France}
\affiliation{Laboratoire de Physique Th\'{e}orique de la Mati\`{e}re Condens\'{e}e, UMR 7600 CNRS, Sorbonne Universit\'{e}, 75005 Paris, France}
\author{Silvia Grigolon}
\email[Electronic address: ]{silvia.grigolon@gmail.com}
\affiliation{Laboratoire Jean Perrin, UMR 8237 CNRS, Sorbonne Universit\'{e}, 75005 Paris, France}
\author{Thibault Bertrand}
\email[Electronic address: ]{t.bertrand@imperial.ac.uk}
\affiliation{Department of Mathematics, Imperial College London, 180 Queen's Gate, London SW7 2AZ, United Kingdom}

\begin{abstract}
\noindent Flocking is a prime example of how robust collective behavior can emerge from simple interaction rules. The flocking transition has been studied extensively since the inception of the original Vicsek model. Here, we introduce a self-propelled particle model with quenched disorder in the pairwise alignment interaction couplings akin to a spin-glass model. We find that increasing the variance of the coupling distribution can promote (rather than destroy) the emergence of global polar order. In particular, we show that our model can display a flocking phase even when the majority of the interaction couplings are antialigning. Activity is the key ingredient to reduce frustration in the system as it allows local particle clustering combined with self-organization of the particles to favor neighborhoods with strong cooperative interactions.
\end{abstract}
\maketitle


How can simple interaction rules lead to robust and unexpected collective behavior? In statistical mechanics, paradigmatic examples are abound, from simple equilibrium spin systems to out-of-equilibrium systems of interacting self-propelled particles (SPP) \cite{ramaswamy2010,marchetti2013,bechinger2016,shaebani2020}. In models of self-propelled particle systems, local consumption of energy can be used to self-organize into ordered states through local alignment interactions alone, leading to striking nonequilibrium behaviors, such as flocking \cite{vicsek1995,toner1995,toner1998,vicsek2012,chate2020}.
Following earlier attempts to model flocking in animal groups \cite{reynolds1987}, Vicsek et al. introduced a model of identical noisy self-propelled particles interacting solely through a local polar alignment rule akin to spins in the XY model \cite{vicsek1995}; this model displays an order-disorder phase transition \cite{vicsek1995,gregoire2004}. Interestingly, despite only having local alignment interaction rules, this model exhibits long-range order in the ordered phase in two dimensions \cite{toner1995,toner1998,tu1998,aditi2002,ramaswamy2003,toner2005,toner2012a,toner2012b,cavagna2015,mahault2019,chate2020}, as activity allows the system to escape the Mermin-Wagner-Hohenberg theorem \cite{mermin1966,hohenberg1967}. Flocking remains under intense scrutiny both theoretically \cite{chate2008a,chate2008b,toner2012a,solon2015} and experimentally including in animal groups \cite{giardina2008,vicsek2012,cavagna2018}, bacterial colonies \cite{peruani2012}, cytoskeletal filaments driven by molecular motors \cite{schaller2010} or synthetic colloidal systems \cite{bricard2013,kaiser2017,geyer2018}. 

Flocking has been shown to be both more robust in some respects and more fragile in others, than originally thought. Indeed, it was recently shown to arise in the absence of explicit alignment interactions and even when these interactions should in fact prevent it \cite{romanczuk2009,deseigne2010,ferrante2013,knezevic2022,casiulis2022,caprini2023,das2024,baconnier2025}. Populations of particles with set inter- and intrapopulation interaction rules can display novel phases not possible in the original Vicsek model \cite{ventejou2021,menzel2012,chatterjee2023,kursten2025,copenhagen2016,ariel2015}. Conversely, although the exact details of the metric alignment rule and implementation of noise in the model do not change the asymptotic properties of the Vicsek model \cite{ginelli2010,ginelli2016,martin2021,martin2024}, the fragility of the ordered state to heterogeneity in SPP models is evident through the influence of obstacles \cite{chepizhko2013,morin2017,codina2022}, spatial anisotropy \cite{solon2022}, or even dissenters \cite{baglietto2013,yllanes2017}. Discrete- and continuous-symmetry flocks with rotational anisotropy have also recently been shown to be metastable \cite{codina2022,Benvegnen2023}. 

Previous works have considered extensions of the continuous-time formulation of the Vicsek model or ``flying XY" model and have studied the influence of density dependent velocities \cite{farrell2012}, chirality \cite{liebchen2017,ventejou2021}, and excluded volume interactions \cite{martin-gomez2018,zhao2021}. Recent studies have focused on the impact of quenched and annealed random field disorder on flocks with homogeneous interaction couplings. Paradoxically, these nonequilibrium systems have been found to be more robust to quenched disorder than the corresponding equilibrium systems \cite{peruani2018,toner2018a, toner2018b,duan2021,chen2022a,chen2022b,chen2022c}.

In equilibrium systems, the role of quenched disorder on phase transitions has been the subject of intense study in the context of models of disordered spins \cite{edwards1975,sherrington1975,binder1986,fischer1991}. Disordered spin systems generically display unresolved frustration arising from the impossibility of satisfying all particle interactions at the same time. This notably leads to the so-called spin-glass phase, characterized by islands of locally aligned spins but no long-range order \cite{mezard1986,fischer1991}. While recent papers have explored geometrical frustration in high-density active systems \cite{paoluzzi2024} and nonreciprocal interactions in XY spin systems \cite{hanai2024}, our understanding of the impact of disordered alignment interactions on the emergence of flocking is still crucially lacking.

Here, we introduce the \textit{active disordered XY} (ADXY) model in which disorder is quenched in the couplings, which are drawn from a Gaussian distribution, as commonly done in equilibrium spin glasses. Quenched disorder is thus introduced at the level of the couplings, a pairwise interaction property rather than a property intrinsic to the particle or the environment. Upon increasing the variance of these spin-spin couplings, we observe nonzero global polar order in systems in which there would otherwise be no long-range order \footnote{Indeed, no long-range order would be observed in a Vicsek model with a non random coupling whose value is set to the same average coupling strength.}. Despite the fact that flocking is disrupted through other forms of coupling heterogeneity, in this case, the variance in the couplings promotes, rather than destroys, flocking; activity leads to a spatial self-organization of the particles that helps decrease frustration in the system.

\paragraph*{Active Disordered XY model.}
We study the dynamics of $N$ pointlike particles interacting solely via metric alignment interactions akin to an off-lattice extension of a two-dimensional (2D) disordered XY model. The particles' dynamics are governed by the following overdamped Langevin equations:
\begin{subequations}
  \begin{align}
    \dot{\rbf}_i &= v_0 \pbf_i, \label{eq:position} \\
    \dot{\theta}_i &= \frac{1}{n_i} \sum_{j\in\mathcal{N}_i} \kij \sin{(\theta_j-\theta_i)} + \eta \xi_i, \label{eq:orientation}
  \end{align}
  \label{eq:langevin}%
\end{subequations}
where $\{\rbf_i, \theta_i\}$ are, respectively, the position and orientation of particle $i\in \{1,\dots,N\}$ at time $t$, $n_i$ is the number of particles in the set of neighbors $\mathcal{N}_i=\{j: |\rbf_i-\rbf_j| \leq \RI, i\neq j\}$, $\eta$ is a constant noise strength, and $\xi_i$ is a Gaussian white noise with zero mean and unit variance. Particles thus move at constant speed $v_0$ along a self-propulsion direction $\pbf_i=(\cos\theta_i, \sin\theta_i)$ that is subject to rotational diffusion with persistence time $\tau_p \propto \eta^{-2}$ and a polar alignment force with couplings $K_{ij}$ from neighboring particles $ j \in \mathcal{N}_i$. Our implementation of the polar forces means that the alignment is nonadditive, which is a more realistic continuous-time version of the original Vicsek model \cite{chepizhko2021}.

 The coupling strength between each particle pair is given by $\kij$, where $\kij>0$ gives ferromagnetic alignment and $\kij<0$ antiferromagnetic alignment. The couplings are independent and identically distributed quenched random variables drawn from a Gaussian distribution $K_{ij}\sim \mathcal{N}(\ka, \ks^2)$ with $\kij=K_{ji}$ as is found in many spin-glass models such as the Sherrington-Kirkpatrick (SK) model \cite{sherrington1975, kirkpatrick1978}. In fact, in the limit of $\RI\to\infty$, we recover a Langevin formulation akin to the SK model, as the spin interactions are infinite ranged and the particle positions in Eq.~(\ref{eq:position}) are irrelevant.

In what follows, we numerically solve Eq.~(\ref{eq:langevin}) in a square box of size $L$ using an Euler-Maruyama method with time step $\Delta t$. Unless stated otherwise, we use periodic boundary conditions, fix $\Delta t = 0.01$, $v_0=1$, and $\RI=1$, and start from initial configurations with random positions and orientations at average density $\rho = N / L^2$. We note that Vicsek-like models are subject to strong finite-size effects; here, we have taken all necessary precautions to ensure that our results are robust, including large numerical simulations of up to order $10^5$ particles.

\paragraph*{Order-disorder transition.}

\begin{figure}
  \includegraphics[width=\linewidth]{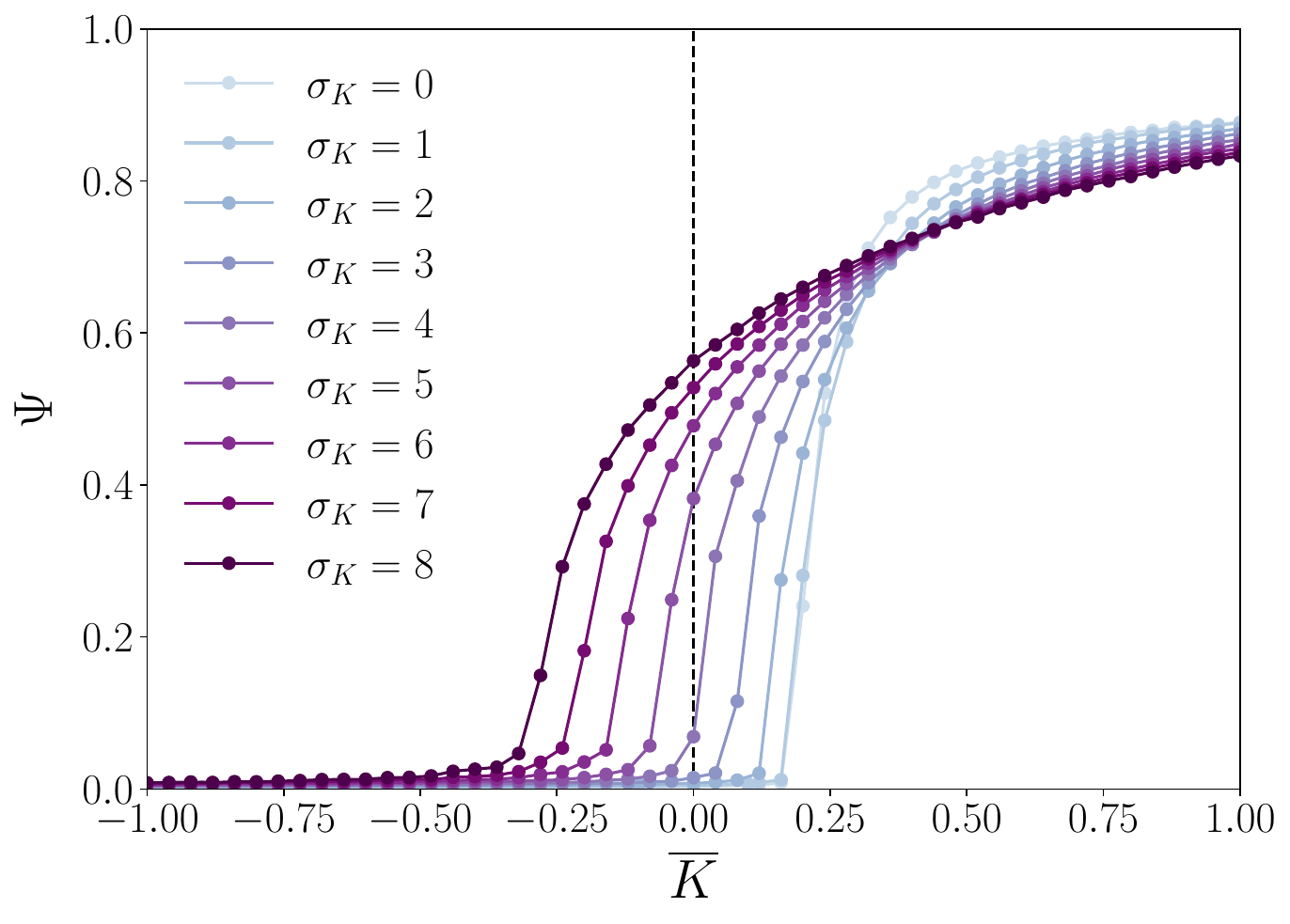}
  \caption{Polar order parameter $\Psi$ vs $\ka$ for increasing values of $\ks$. Simulations were performed with $N=9\times10^4, \ \rho=1, \ \eta=0.4$. The black dashed line denotes $\ka=0$.}
  \label{fig:phase_transition}
\end{figure}

\begin{figure*}
  \includegraphics[width=\linewidth]{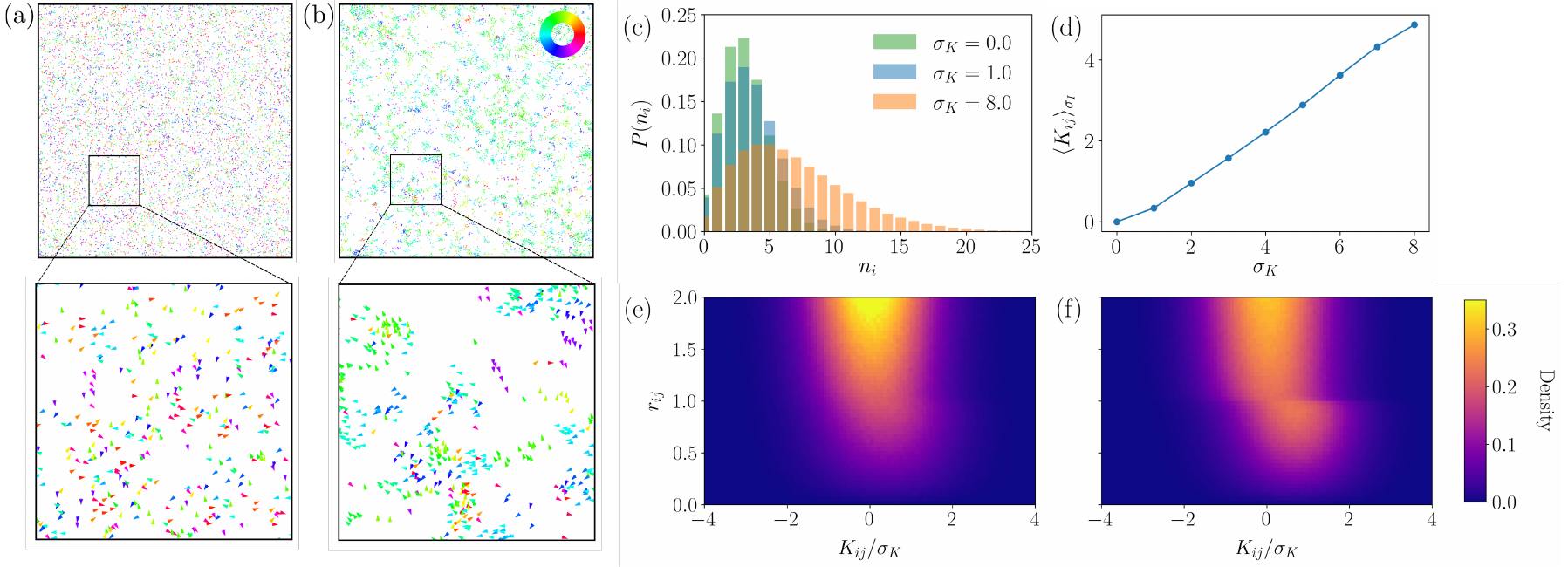}
  \caption{Local clustering of particles. Snapshots from simulations with $\ks=1$ in panel (a) and $\ks=8$ in panel (b), with $N=10^4, \ \rho=1, \ \eta=0.4$, and $\ka=0$. Close-up views of the simulation box are shown below with length $L/5=20$. The colored arrows give the particle orientations $\theta\in[0,2\pi)$. See movies S1 and S2 and their descriptions in the Supplemental Material \cite{Note2}. (c) Normalized histograms for the number of neighbors $n_i$ within the radius of interaction $\RI=1$ for the systems in (a) and (b) with $\ks=1$ in blue and $\ks=8$ in orange, as well as $\ks = 0$ (noninteracting particles) in green. (d) Mean $\kij$ over interacting pairs in steady state against $\ks$ for $\ka=0$. 2D histograms of distance between particle pairs $r_{ij}=|\rbf_i-\rbf_j|$ against their coupling value for $\ks=1.0$ [panel (e)] and $\ks=8.0$ [panel (f)]. Coupling values have been normalized by $\ks$. Data are averaged over time (at intervals of $t=100$ from $t=3000$ to $3500$), and realizations, with $N=10^4, \ \rho=1, \ \eta=0.4$, and $\ka=0$. Note that histograms remain symmetric for values of $r_{ij} > 2$, not shown here.}
  \label{fig:snaps_hist}
\end{figure*}

The degree of global polar order in a flocking system is measured through the time-averaged global polar order parameter $\Psi=\langle\psi(t)\rangle_t$, where
\begin{equation}
  \psi(t) = \frac{1}{N} \left| \sum_{i=1}^N \pbf_i(t)\right|,
\end{equation}
where $\Psi$ takes values from $\psi=0$ (disorder) to $\psi=1$ (order). Averages are taken over time (after the system reaches steady state) as well as realizations of the random couplings $\kij$ (usually 20--50 realizations).

For homogeneous couplings---$\ks=0$ and $\kij=\ka$ for all pairs $(i,j)$---a disorder-to-order phase transition is observed as the value of $\ka$ is increased \cite{gregoire2003,gregoire2004,farrell2012,sese-sansa2018,martin-gomez2018}. We confirm this result here. For $\ks>0$, we also observe such a transition (see Fig.\,\ref{fig:phase_transition}). However, as $\ks$ is increased, the onset of flocking strikingly shifts to negative values of $\ka$. In the region $\ka\leq 0$, order is promoted as $\ks$ increases, and we are able to recover a significant degree of global order, despite the fact that the majority of interactions are \emph{antialigning}. In the case $\sigma_K=0$, recent studies argue that Vicsek-type models with metric-free interactions behave as do their metric counterpart \cite{martin2021,martin2024}. Here, we confirm that our observations are robust and extend to systems with non-Gaussian distributions of random couplings and systems with topological (rather than metric) interactions (see details in Supplemental Material \cite{Note2}). 

To understand the emergence of this mechanism, we compare snapshots from simulations with small and large $\ks$ values in Fig.\,\ref{fig:snaps_hist} (see Supplemental Material \cite{Note2} for movies). To start, we consider systems with $\ka=0$, in which there are an equal number of aligning and antialigning interactions.
At $\ka=0$ and $\ks=1$, the polar order parameter is $\Psi=0.01$: The system is disordered and particles point in random directions [Fig.\,\ref{fig:snaps_hist}(a)]. Upon increasing the coupling dispersion to $\ks=8$, the steady-state polar order parameter increases to $\Psi=0.57$. This increase in the polar order parameter corresponds to the appearance of clusters of locally aligned particles [Fig.\,\ref{fig:snaps_hist}(b)]. Indeed, as $\ks$ is increased, the distribution of number of neighbors $n_i$ within the interaction radius becomes more right skewed and particles are more likely to have a greater number of neighbors, as shown in Fig.\,\ref{fig:snaps_hist}(c).

To investigate this clustering further, we look at the relationship between particle pair rescaled couplings $\kij/\ks$ and their separation distance $r_{ij}$ in Fig.\,\ref{fig:snaps_hist}(e)--(f). Owing to the Gaussian nature of the coupling distribution, we would expect the joint probability distribution $P(K_{ij}/\ks,r_{ij})$ to be peaked around $K_{ij}/\ks =0$ and fall off symmetrically if particles were clustered randomly. This is what we observe for distances $r_{ij} > \RI$. However, for $r_{ij}<\RI=1$, we observe a clear shift toward positive coupling values, which becomes more pronounced as $\ks$ increases. In other words, an increase in $\sigma_K$ promotes the creation of more cohesive clusters of highly aligning particles. Indeed, we find that the mean coupling within the interaction radius $\langle\kij\rangle_{\RI}$ increases monotonically with $\ks$, as shown in Fig.~\ref{fig:snaps_hist}(d).
While the variance in the couplings introduces an inherent frustration in the particle orientations, the activity allows particles to self-organize to find other particles to satisfy their interactions, thus decreasing the global frustration in the system. However, this does not imply that all antiferromagnetic couplings ($\kij<0$) become irrelevant. Even at large $\ks$, a significant proportion of interactions remain frustrated, in the sense that they contribute unfavorably to the energy (see Supplemental Material \cite{Note2}).

\begin{figure}
    \centering
\includegraphics[width=\linewidth]{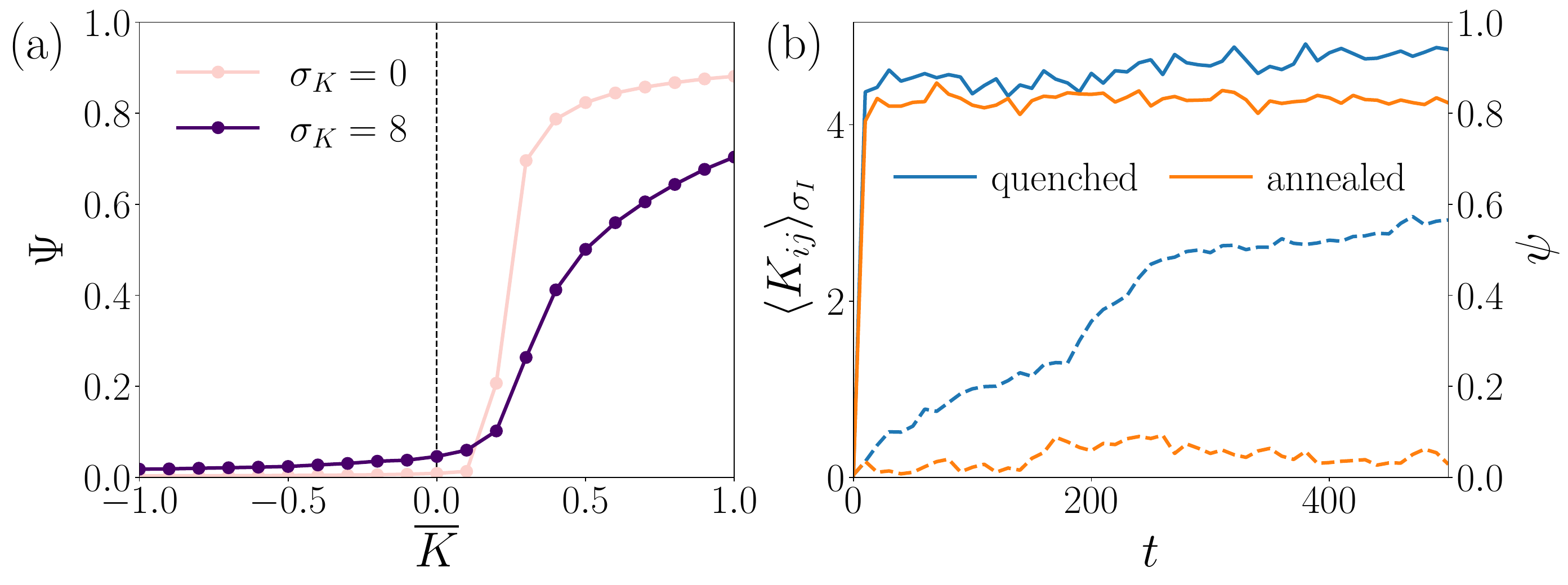}
\caption{(a) Polar order phase transition against $\ka$ for the annealed variant of coupling disorder. Simulation parameters were $N=10^4$, $\rho=1$, and $\eta=0.4$. (b) Comparison of quenched and annealed coupling disorder for one realization with $\ka=0$, $\ks=8$. Mean $\kij$ over interacting pairs over time (solid lines) and polar order $\psi$ over time (dashed lines).}
    \label{fig:annealed}
\end{figure}

\begin{figure*}
  \includegraphics[width=\linewidth]{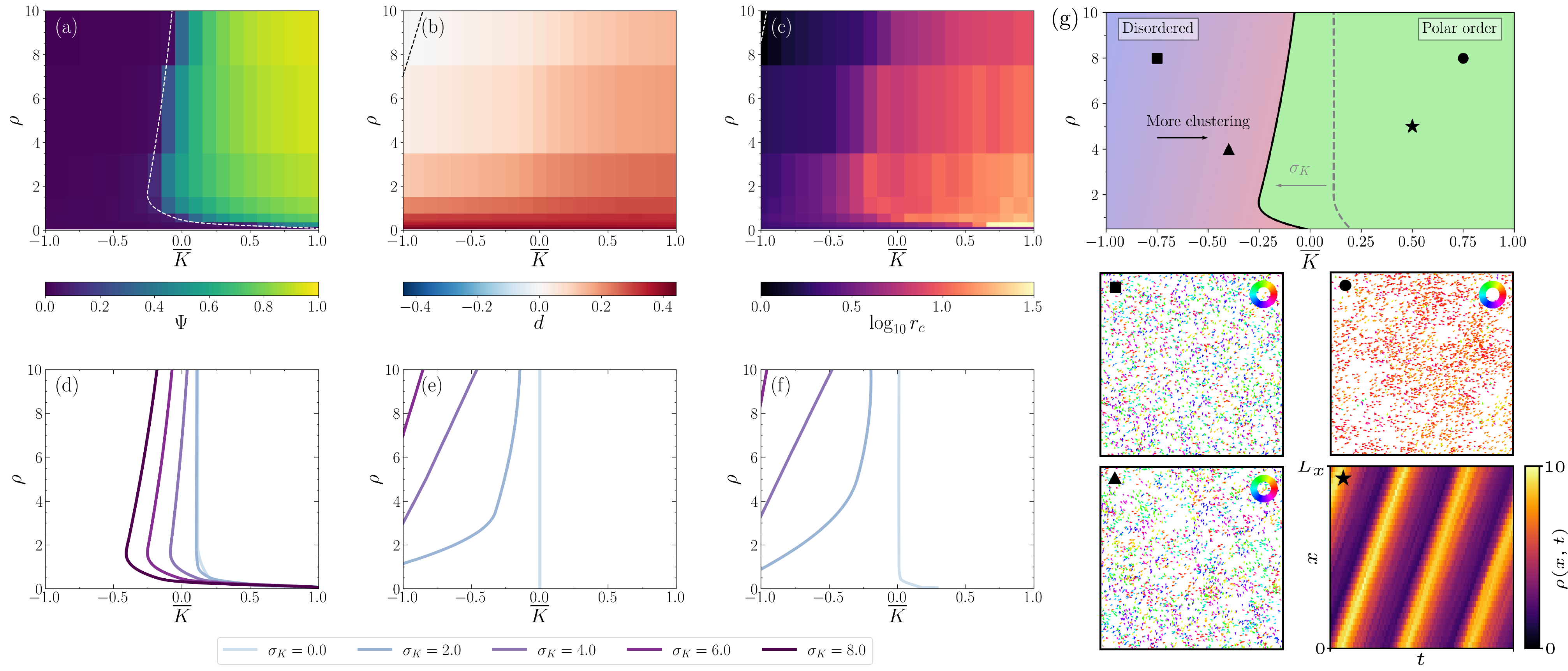}
  \caption{Phase diagrams in the $(\ka, \rho)$ plane. (a) Color map of the polar order parameter with a dashed line showing the order-disorder transition at $\Psi=0.1$. (b) Color map of the demixing order parameter with a dashed line for the transition at $d=0$. (c) Color map of the clustering order parameter, with a dashed line at $r_c=1$. For panels (a)--(c), the parameters are as follows: $N=10^4$, $\eta=0.4$, and $\ks=6$. (d)--(f) Transition lines for various values of $\ks$ for the corresponding polar, demixing, and clustering order parameters shown above. Transition lines for a fixed value of $\ks = 6$ are plotted on the same graph in Supplemental Material \cite{Note2}. (g) Phase diagram for $\eta=0.4$ and $\ks=6$ based on the phases determined by the three order parameters. The gray dashed line shows the polar order transition line for $\sigma_K=0$ (i.e., no coupling disorder). This line shifts toward the left as $\sigma_K$ increases. Example snapshots from simulations are shown below (zoomed in to $L/2\times L/2$) for $\ka=-0.75$, $\rho=8$ (square); $\ka=-0.4$, $\rho=4$ (triangle); $\ka=0.75$, $\rho=8$ (circle). Additionally, a kymograph of the density transverse to the band's direction against time is plotted for $\ka=0.5$, $\rho=5$ (star) with $N=10^5$, $L_x/L_y=5$ to show the presence of banding in the polar ordered phase. See movies S3--S6 and their descriptions in the Supplemental Material \cite{Note2}.}
  \label{fig:phase_diagrams_snaps}
\end{figure*}

The formation of highly aligned clusters alone does not necessarily imply global polar order. We simulated an annealed variant in which $\kij$ remains fixed while particles $i$ and $j$ are neighbors, but is updated if they separate and later come back within each other's interaction radius. This update rule ought to favor local clusters of highly aligning particles. However, in Fig.~\ref{fig:annealed}, we clearly observe a loss of global cooperation when the coupling distribution is replaced with this type of annealed disorder. For $\ka<0$, increasing the coupling variance no longer leads to global polar order [Fig.~\ref{fig:annealed}(a)]. Comparing a single realization of the quenched and annealed model variants in Fig.~\ref{fig:annealed}(b), the mean $\kij$ over interacting pairs rapidly increases to a similar level in both cases, signifying highly aligned clusters. However, while the global order of the quenched disorder simulation steadily increases over time to a steady state of $\psi_\infty\approx 0.6$, the simulation with the annealed disorder variant remains globally disordered. This suggests that quenched disorder generates an effective interaction network capable of sustaining alignment beyond individual clusters. Recently, a nonreciprocal variant of this model was also observed to show strong clustering, yet no global polar order \cite{choi2025}. Taken together, these results suggest that the key ingredients for global ordering in the ADXY model with $\ka<0$ and $\ks>0$ are activity and reciprocal couplings with quenched disorder.

\paragraph*{Self-organization and clustering promote polar order.}  
To systematically characterize the phases of our model, we first measure the polar order parameter $\Psi$ for a fixed noise strength $\eta=0.4$ while varying $\rho$, $\ka$, and $\ks$ [Fig.\,\ref{fig:phase_diagrams_snaps}(a) and (d)]. Remarkably, for large enough values of $\ks$, the system exhibits a region of parameter space with global polar order but $\ka \leq 0$; this is made possible thanks to the self-sorting mechanism discussed above. Note that as the density $\rho$ is varied, we observe reentrance in the polar order parameter not seen for homogeneous couplings ($\ks =0$); indeed, the observed critical value $\ka_c$ for the onset of flocking ($\Psi >0$) varies nonmonotonically with the density and displays a maximum around $\rho \approx 2$ for our choice of parameters.

As $\rho \to 0$, we expect a transition to disorder, as seen in the original Vicsek model \cite{vicsek1995}. At low density, the lack of interaction prevents the propagation of information. At very high densities, particles have many interacting neighbors and the self-sorting clustering mechanism needed for polar order to arise is impaired due to a greater frustration in the system; the number of interacting neighbors outweighs the number of large, positive couplings $\kij$, thus destroying global polar order.

To further support self-organization and clustering as a mechanism to promote polar order when $\ka<0$, we define the demixing order parameter as
\begin{equation}
  d = \left\langle \frac{1}{N}\sum_{i=1}^N \frac{\kappa_i}{n_i} -\frac{1}{2} \right\rangle
\end{equation}
where $n_i$ is the total number of interacting neighbors of particle $i$ within the radius of interaction $\sigma_I$ and $\kappa_i$ is the number of interacting neighbors with $\kij>0$. Their ratio therefore quantifies the proportion of ferromagnetically interacting neighbors, meaning that $d=0$ corresponds to a perfectly mixed system and $d>0$ (resp. $d<0$) to a system with more local ferromagnetic (resp. antiferromagnetic) interactions on average. The average $\langle\cdot\rangle$ is taken over time and realizations of $\kij$, removing particles with no neighbors from the sum.

In Fig.\,\ref{fig:phase_diagrams_snaps}(b), we show the value of $d$ in the $(\ka,\rho)$ plane and draw a transition line at $d=0$ (see Fig.\,\ref{fig:phase_diagrams_snaps}(e) for the transition lines for several values of $\ks$). For completely mixed systems, we would expect a vertical transition line at $\ka=0$. However, we notably find regions of parameter space where $d > 0$ despite $\ka < 0$, in line with the clustering mechanism described above and seen in Fig.\,\ref{fig:snaps_hist}(e).
We also note that a heuristic mean-field calculation yields a minimal nucleation criterion for the appearance of rare cooperative subsets, confirming the presence of cooperative nuclei with positive mean coupling value when $\ka<0$ and $\ks$ is sufficiently large (see Supplemental Material Ref.~\cite{Note2} for details). The active dynamics can subsequently amplify such cooperative subsets into global order, although this lies beyond the mean-field description.

For each system, we also measure the normalized equal-time connected density correlations $C_\rho(r)$ (see Supplemental Material \cite{Note2}). We define the typical cluster size $r_c$ to be the radial distance such that $C_\rho(r_c) = 0.01$. In a homogeneous system, we would expect correlations to decay very quickly outside the radius of interaction, leading to $r_c\approx\sigma_I$. For systems with clustering, density fluctuations increase such that $r_c\gg\sigma_I$. Typical cluster sizes are reported on a log scale in Fig.\,\ref{fig:phase_diagrams_snaps}(c) for $\ks=6$. We define the boundary between homogeneous and clustered systems as $r_c=\sigma_I$. Note that the original Vicsek model ($\ks = 0$) also displays clustering in the disordered phase close to the transition \cite{chakraborti2019}, though this region is much smaller in the $(\ka, \rho)$ space compared to the ADXY model. Comparing Figs.\,\ref{fig:phase_diagrams_snaps}(d)--(f), we find that the transitions for demixing and clustering coincide with each other. Systems in this clustered disordered region display local ordering within highly aligning clusters but have not become clustered enough to propagate this information far enough to display global polar order. Similar disordered clustered states were found in another active particle model with distance dependent alignment \cite{grossmann2015}, and more recently in an ADXY model with nonreciprocal alignment \cite{choi2025}. Finally, note that the clustering transition observed in the disordered phase is accompanied by a transition from hyperuniform to clustered states (see Supplemental Material \cite{Note2}).

\paragraph*{Phase diagram for the active disordered XY model.}  
We propose a phase diagram for the ADXY model in Fig.\,\ref{fig:phase_diagrams_snaps}(g). For $\ka\ll0$, we have the familiar disordered homogeneous phase. As $\ka$ increases, clustering sets in. Once $\ka$ is increased further, we reach the polar ordered flocking phase, which includes regions of parameters space where $\ka\leq0$. Note that the onset of polar order is accompanied by a sharp increase in the particles' persistence and the typical pairwise contact time (see Supplemental Material \cite{Note2}). Polar order decreases monotonically with increasing noise strength $\eta$, as in the Vicsek model (see Supplemental Material \cite{Note2}).

\footnotetext{Toner and Tu predicted the connected, equal-time, velocity autocorrelation to scale $C(r)\propto r^{2\chi/\zeta}$ for sufficiently large $r$, with $\chi=-1/5$, $\zeta=3/5$ in $d=2$ \cite{toner1998}. These scaling exponents were later shown not to be exact \cite{toner2012a}; however, recent work still shows the same scaling of $\omega=2\chi/\zeta=-2/3$ \cite{jentsch2024, chate2024}. This exponent was found to be approximately $-0.65$ by others numerically \cite{mahault2019}.}

\begin{figure}[t]
  \includegraphics[width=\linewidth]{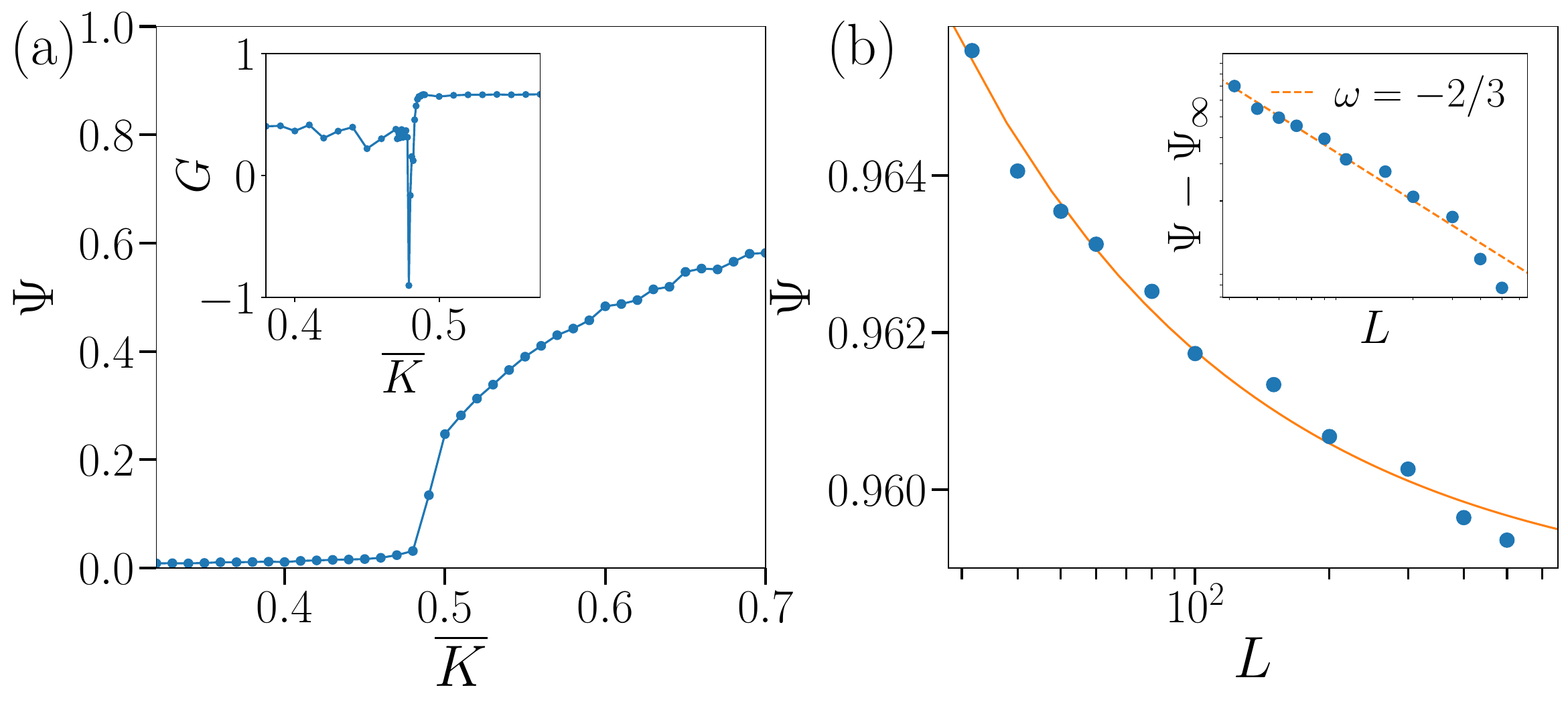}
  \caption{Nature of the phase transition and ordered phase. (a) Polar order parameter $\Psi$ vs $\ka$ with $\ks=1$. Inset: Binder cumulant $G$ vs $\ka$ showing a negative dip around the transition point. Simulations were performed with $N=6\times10^4$, $\rho=1$, and $\eta=0.6$. (b) Polar order parameter $\Psi$ vs system size $L$ with a line showing a fitted curve, which has algebraic decay to an asymptotic value of $\Psi_\infty=0.958(5)$, fitted with a scaling exponent of $\omega=-2/3$. The inset shows the log-log fit of $\Psi-\Psi_\infty\sim L^{\omega}$ in a dashed line for the data (blue dots). Simulations were performed with $\rho=1, \ \eta=0.2, \ \ka=1, \ \ks=1$.}
  \label{fig:binder_LRO}
\end{figure}

We also find a banding phase between the disordered phase and the homogeneous polar liquid phase (or Toner-Tu phase). For $N=10^4$, this manifests as clustered systems with large density fluctuations due to finite-size effects, as shown in Fig.\,\ref{fig:snaps_hist}(b). However, for large enough system sizes, we recover ordered traveling bands, as reported for the original Vicsek model \cite{chate2008b}. We found our model to be subject to even stronger finite-size effects, requiring for instance, even larger system sizes and a slab geometry to observe stable banding. A kymograph of the local density in slices transverse to the direction of travel reveals clear and stable periodic bands moving in time, with the largest density close to the front of the band, leaving a density trail in its wake (see details in Supplemental Material \cite{Note2}).
The presence of bands, together with a negative dip in the Binder cumulant $G = 1- {\langle\psi^4\rangle}/{3\langle\psi^2\rangle^2}$ near the transition for small $\ks>0$ [Fig.~\ref{fig:binder_LRO}(a)], supports a first-order order-disorder transition, as in the original Vicsek model \cite{gregoire2004,chate2020}. Although for larger $\ks$ we could not clearly resolve this dip within the accessible system sizes, our results suggest that quenched disorder in the coupling strengths does not change the nature of the transition. 
We further find evidence that the polar liquid phase retains true long-range order: For $\ks=1$, the global polar order is consistent with an algebraic decay to a nonzero asymptotic value $\Psi_\infty$ and its system-size dependence is well described by $\Psi-\Psi_\infty\sim L^{-2/3}$, consistent with the predictions of Toner and Tu \cite{Note3}.

Finally, we observe that the flocks obtained for $\ks>0$ appear to be more fragile than those for $\ks = 0$; indeed, the microscopic angular diffusion coefficient in the flocking phase is larger when quenched disorder is present at all values of $\ka$ (see Supplemental Material \cite{Note2}).

\paragraph*{Conclusion.}
In a swath of natural (including cellular tissues, bacterial suspensions, and bird flocks) and synthetic (including active colloids, self-propelled bots, cell colonies, and organoids) systems, agent-to-agent variability is the norm \cite{ariel2022,topaz2012,zuo2020,peled2021,bertrand2024}, and we expect randomly distributed alignment interactions to be the rule rather than the exception.
Inspired by this, we introduced the ADXY model as an extension of existing models of self-propelled interacting particles to study the consequences of random alignment interaction for collective motion. Our main finding is that, when combined with activity, such coupling disorder can provide a general mechanism for global order, even for \emph{negative} values of $\ka$. In turn, we show that increasing the variance in the couplings can promote rather than destroy flocking, a surprising result when compared to recent studies \cite{baglietto2013,yllanes2017}. Importantly, we show that this counterintuitive phenomenon is explained by a self-organization mechanism in which activity allows particles to rearrange into local neighborhoods with favorable alignment couplings, thereby reducing the local frustration present in a spatially uniform configuration. Our results further show that local clustering alone is not sufficient for global flocking, pointing instead to the crucial role of quenched reciprocal couplings in connecting local alignment to system-spanning order.
In particular, we showed that the system exhibits clustering even before the onset of collective motion. Future work will attempt to explore the interplay between activity and frustration, and examine whether a spin-glass phase can exist in our model and more generally in active systems.
Addressing these questions is analytically challenging and may benefit from combining analytical techniques from both active matter and spin-glass theory.
While the work presented here is theoretical, our findings may offer a new perspective on the roles of heterogeneity and variability in natural swarms and cell-cell interactions.

\paragraph*{Acknowledgments.}
We thank Hugues Chat\'e, Silvio Franz, Ananyo Maitra, Alessandro Manacorda, Enzo Marinari, Roberto Menichetti, and Francesco Zamponi for useful discussions. E.L. was funded by a President's Ph.D. Scholarship at Imperial College London. S.G. acknowledges the ICL-CNRS fellowship and the grant Tremplin@INP by CNRS Physics as well as the hospitality of the Department of Mathematics at the Imperial College London. The authors acknowledge computing resources provided by Imperial College Research Computing Service and the Department of Mathematics Compute Cluster.

\paragraph*{Data availability.} The data that support the findings of this article are openly available \cite{lardet2026}.

\end{document}


\title{Supplemental Material for ``Disordered Yet Directed: The Emergence of Polar Flocks with Disordered Interactions"}
\date{\today}

\author{Eloise Lardet}
\affiliation{Department of Mathematics, Imperial College London, 180 Queen's Gate, London SW7 2AZ, United Kingdom}
\author{Rapha\"el Voituriez}
\affiliation{Laboratoire Jean Perrin, UMR 8237 CNRS, Sorbonne Universit\'{e}, 75005 Paris, France}
\affiliation{Laboratoire de Physique Th\'{e}orique de la Mati\`{e}re Condens\'{e}e, UMR 7600 CNRS, Sorbonne Universit\'{e}, 75005 Paris, France}
\author{Silvia Grigolon}
\email[Electronic address: ]{silvia.grigolon@gmail.com}
\affiliation{Laboratoire Jean Perrin, UMR 8237 CNRS, Sorbonne Universit\'{e}, 75005 Paris, France}
\author{Thibault Bertrand}
\email[Electronic address: ]{t.bertrand@imperial.ac.uk}
\affiliation{Department of Mathematics, Imperial College London, 180 Queen's Gate, London SW7 2AZ, United Kingdom}

\maketitle

\hrule height 0.5mm

\begingroup
\hypersetup{linkcolor=black}
\tableofcontents
\endgroup

\vspace{2em}
\hrule height 0.5mm

\section{Supplementary movies}
We provide supplementary movies accompanying Fig.~2 and Fig.~4 of the main text. Particle orientation polar histograms are included alongside the particle trajectories for Movies S1-S5. Movie S6 shows an example movie of banding.

The parameters used are given as follows:
\begin{itemize}
  \setlength\itemsep{0em}
  \item Movie S1: $N=10^4$, $\rho=1$, $\eta=0.4$, $\ka=0$, and $\ks=1$.
  \item Movie S2: $N=10^4, \ \rho=1, \ \eta=0.4, \ \ka=0$, and $\ks=8$.
  \item Movie S3: $N=10^4, \ \rho=8, \ \eta=0.4, \ \ka=-0.75$, and $\ks=6$.
  \item Movie S4: $N=10^4, \ \rho=4, \ \eta=0.4, \ \ka=-0.4$, and $\ks=6$.
  \item Movie S5: $N=10^4, \ \rho=8, \ \eta=0.4, \ \ka=0.75$, and $\ks=6$.
  \item Movie S6: $N=10^5, \ \rho=5, \ \eta=0.4, \ \ka=0.5$, and $\ks=6$, in a slab geometry with $L_x/L_y=5$.
\end{itemize}

\section{Confirming the phase transition}

\begin{figure}[h!]
  \includegraphics[width=\linewidth]{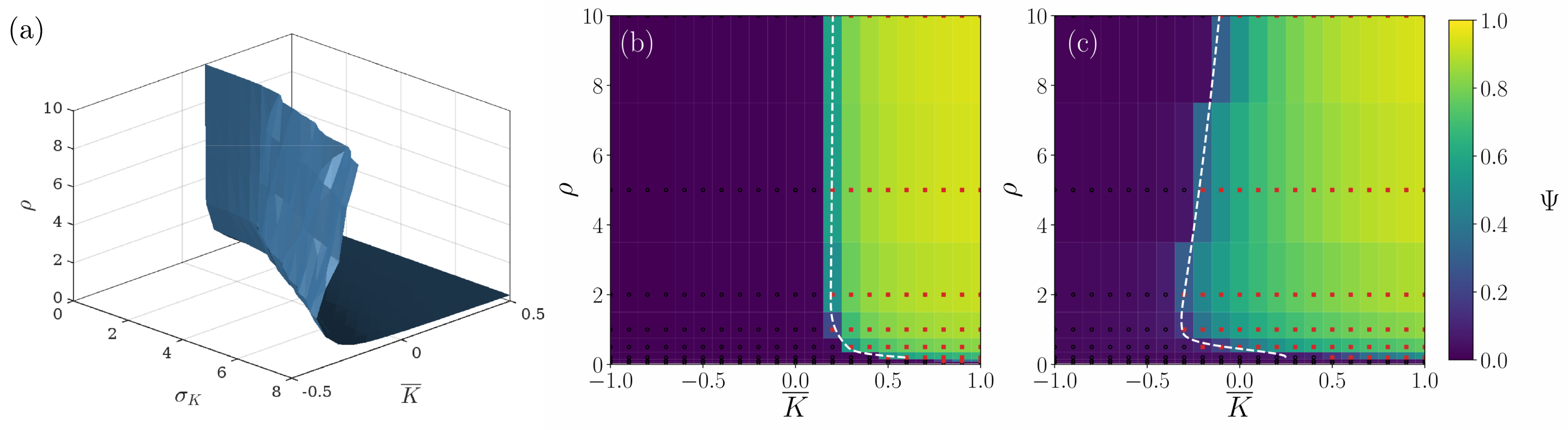}
  \caption{Order-disorder phase transition in the ADXY model. (a) Surface plot for boundary between disordered and ordered polar phase. (b)-(c) 2d phase plot slices of (a) for $\ks=0$ (b) and $\ks=8$ (c). The white dashed lines show the boundaries between the disordered and ordered polar phase. Simulations were performed with $N=10^4, \ \eta=0.4$.}
  \label{fig:3d_phase_diagram}
\end{figure}

In the main text, we provided phase diagrams for the ADXY model with normally distributed interaction couplings and metric interactions. Here, we further provide in Fig.~\ref{fig:3d_phase_diagram} a surface plot of the order-disorder transition in the space $(\ka,\ks,\rho)$ for $\eta=0.4$. We clearly see the presence of re-entrant behavior with increasing $\ks$.

We also combine in a single phase diagram the three transition lines described in Fig.~4 of the main text (for a slightly lower value of $\ks$). 
\begin{figure}[h!]
  \includegraphics[width=0.5\linewidth]{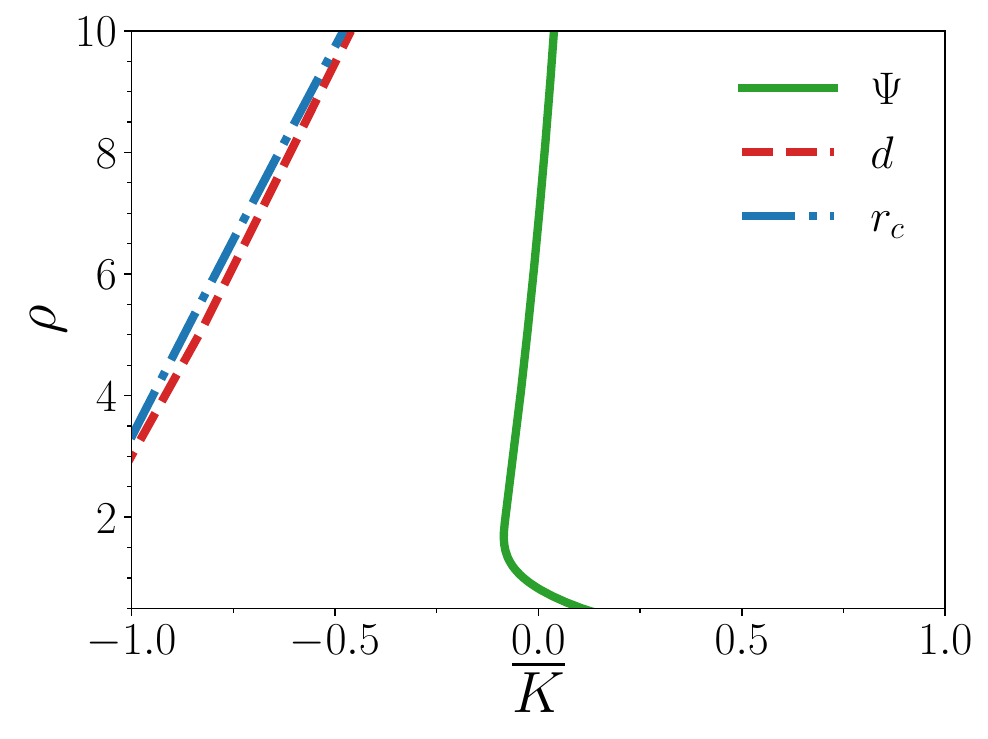}
  \caption{Transition lines in the ADXY model for a single value of $\ks = 4$ including the transition lines defined through the global order parameter $\Psi$, the demixing order parameter $d$ and the typical cluster size $r_c$ as measured using local density correlation functions. Simulations were performed with $N=10^4, \ \eta=0.4$.}
  \label{fig:all_lines_phase_diagram}
\end{figure}

\begin{figure}[h!]
    \centering
    \includegraphics[width=0.5\linewidth]{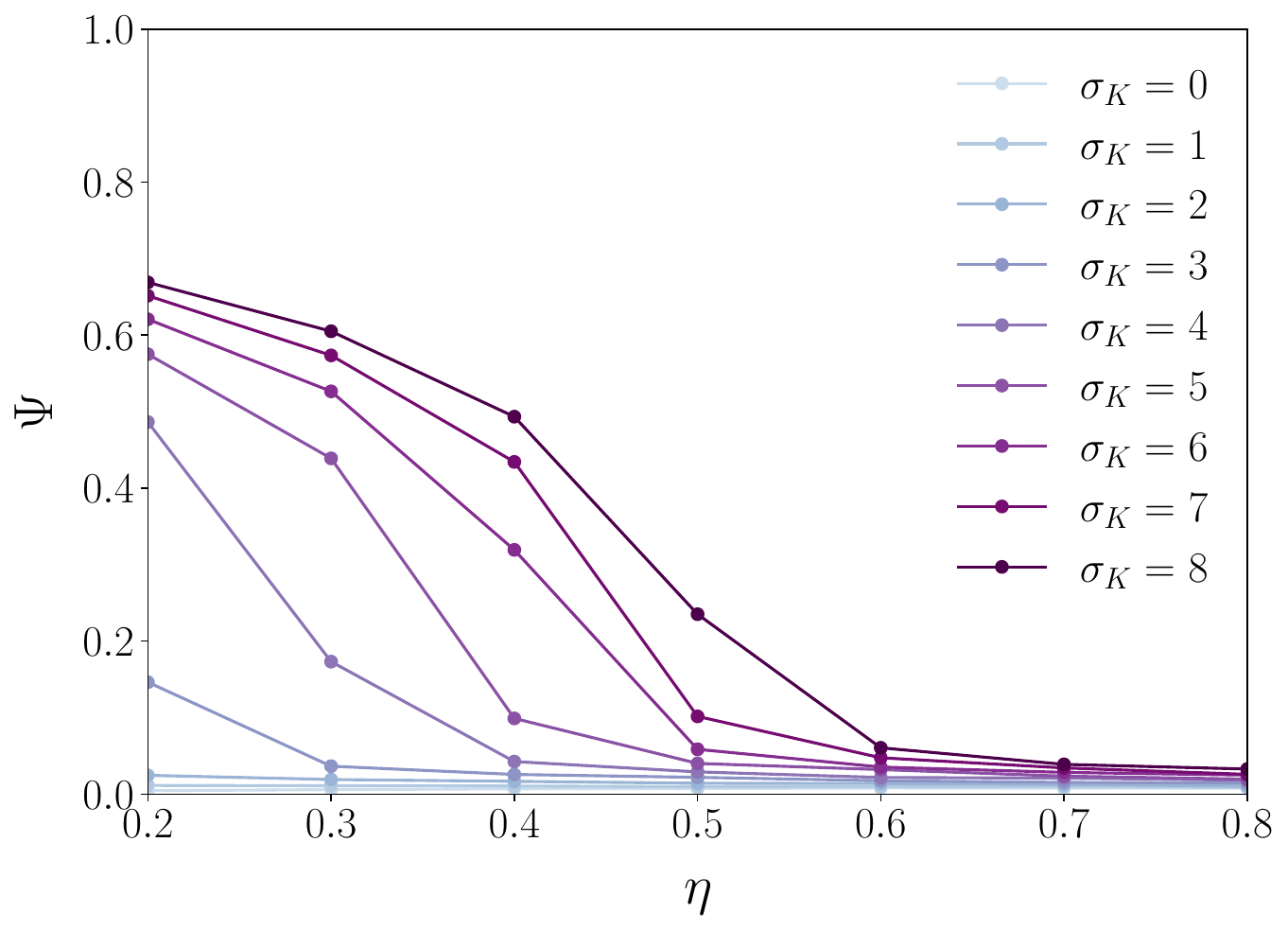}
    \caption{Polar order parameter $\Psi$ against noise strength $\eta$ for different $\ks$, at fixed $\ka=-0.1$.}
    \label{fig:psi_vs_noise}
\end{figure}

We also briefly study how the global polar order varies as we increase the angular noise, keeping all other parameters constant. The global order simply decreases monotonically as noise strength increases, for all values of $\ka$ and $\ks$ we tested (see Fig.\,\ref{fig:psi_vs_noise} below for the $\ka=-0.1$ case), just as in the original Vicsek model. To reduce the parameter exploration, we thus focus in the main text on exploring the phase diagram of this model only in terms of $\ka$, $\ks$ and $\rho$.

To confirm their robustness, we also confirmed that our results remain qualitatively unchanged for a different interaction coupling distribution and if topological interactions are used instead of metric ones. We also consider the case of annealed, instead of quenched, interaction couplings.

\subsection{Weighted bimodal coupling distribution}

First, we modify the coupling distribution while maintaining the first two moments of the original Gaussian coupling distribution. We introduce a weighted bimodal coupling distribution such that couplings only take one of two distinct values. Specifically, we consider a probability distribution with two modes $K_+$ and $K_-$, such that $K_+$ is chosen with probability $\alpha$. 
We can easily write down the mean and standard deviation of this distribution:
\begin{align}
\ka &= \alpha K_+ + (1-\alpha)K_-, \\
\ks &= \left|K_+ - K_-\right|\sqrt{\alpha(1-\alpha)}.
\end{align}
If we fix $\ka, \ks$, and $\alpha$, it is simple to calculate the values of $K_+$ and $K_-$:
\begin{align}
  K_+ &= \ka + \ks\sqrt{\frac{1-\alpha}{\alpha}}, \\
  K_- &= \ka - \ks\sqrt{\frac{\alpha}{1-\alpha}}.
\end{align}
To relate this distribution to a Gaussian distribution with the same $\ka$ and $\ks$, we can choose $\alpha$ such that this is equal to the probability that $\kij>0$ in the Gaussian distribution $\mathcal{N}(\ka,\ks)$, which can be calculated through a simple integration:
\begin{equation}
  \alpha \equiv \int_0^\infty \frac{1}{\ks \sqrt{2\pi}}e^{-\frac{1}{2}\left(\frac{x-\ka}{\ks}\right)^2} \ dx.
\end{equation}
Implementing this distribution in our model with reciprocal couplings, we show the phase transitions in Fig.~\ref{fig:F_top_pt} as a function of $\ka$ for various values of $\ks$. As $\ks$ is increased, we see the same shift of the onset of flocking to negative $\ka$ as in the Gaussian coupling distribution model as in Fig.~\ref{fig:3d_phase_diagram}. 

\subsection{Topological interactions}

\begin{figure}
  \includegraphics[width=\linewidth]{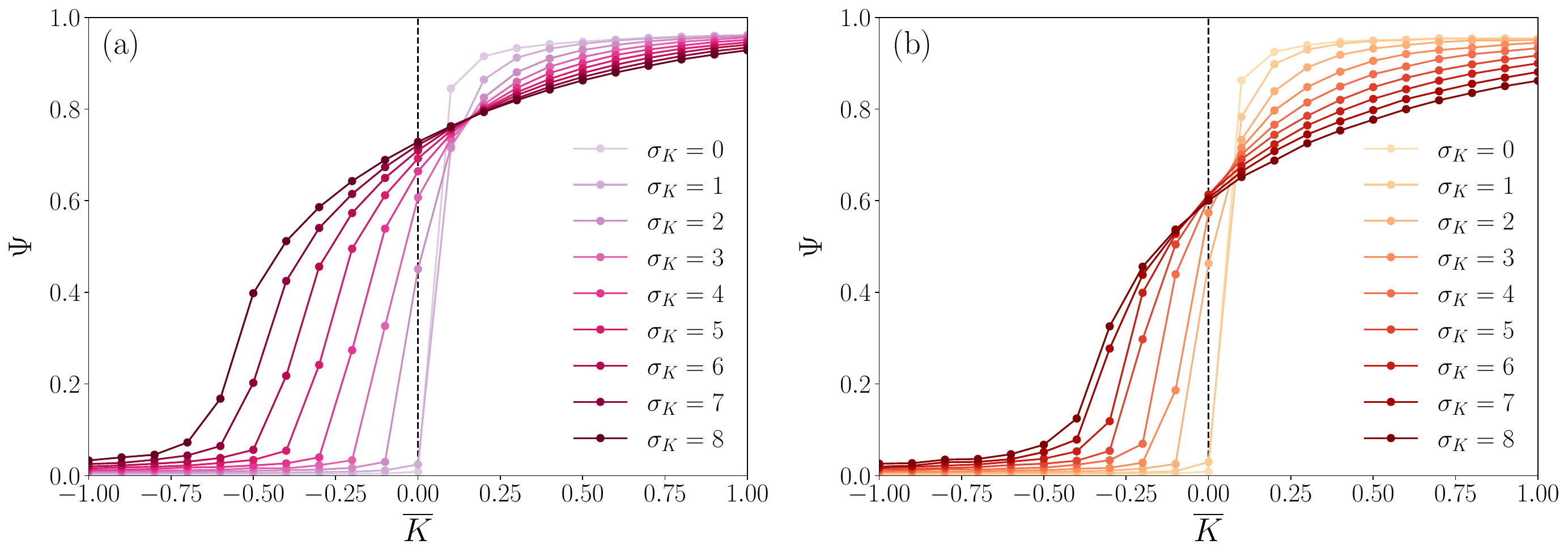}
  \caption{Polar order parameter $\Psi$ vs $\ka$ for increasing $\ks$ with (a) weighted bimodal couplings or (b) topological interactions. Simulations were performed with $N=10^4, \ \rho=1$,  and $\eta=0.2$. The interaction range is $\RI=1$ in (a) and $k=7$ neighbors in (b). The black dashed lines show the vertical line $\ka=0$.}
  \label{fig:F_top_pt}
\end{figure}

Secondly, we use topological interactions with $k$ nearest neighbors, rather than metric interaction range as in the main text. To do so, we simply redefine the interaction neighbors of each particle $\mathcal{N}_i$ to be the $k$-nearest neighbors, in terms of the $L_2$ norm. We show the phase transition for this case in Fig.~\ref{fig:F_top_pt}(b) as a function of $\ka$ as $\ks$ increases. Again, we clearly see the same shift of the onset of flocking to negative $\ka$ as $\ks$ increases, just as in the main text.

\subsection{Annealed dynamics}

In the study of spin glasses, the presence of quenched disorder---where interactions between particles are fixed and non-fluctuating---makes analytical calculations notoriously difficult and generically leads to the rugged potential energy landscape and frustration. To simplify matters, a common but less realistic approximation is to consider annealed couplings; under this approximation, couplings are allowed to dynamically evolve over timescales comparable to the spin dynamics. In these systems, averages are taken at the level of the partition function rather than the free energy as a result the disorder is averaged out, effectively eliminating frustration from the system. We claim that this is not the case in our model, which, like the Edwards-Anderson (EA) model, is defined by its quenched disorder. Indeed, the self-organization and clustering of particles with strong, positive interactions do reduce some frustration and the active dynamics of the particles is sufficient to promote global polar order. This emergence of global polar order thus arises from the unique interplay of quenched disorder and activity, a mechanism that an annealed approximation would not capture.

\begin{figure*}[t!]
    \centering
\includegraphics[width=0.5\linewidth]{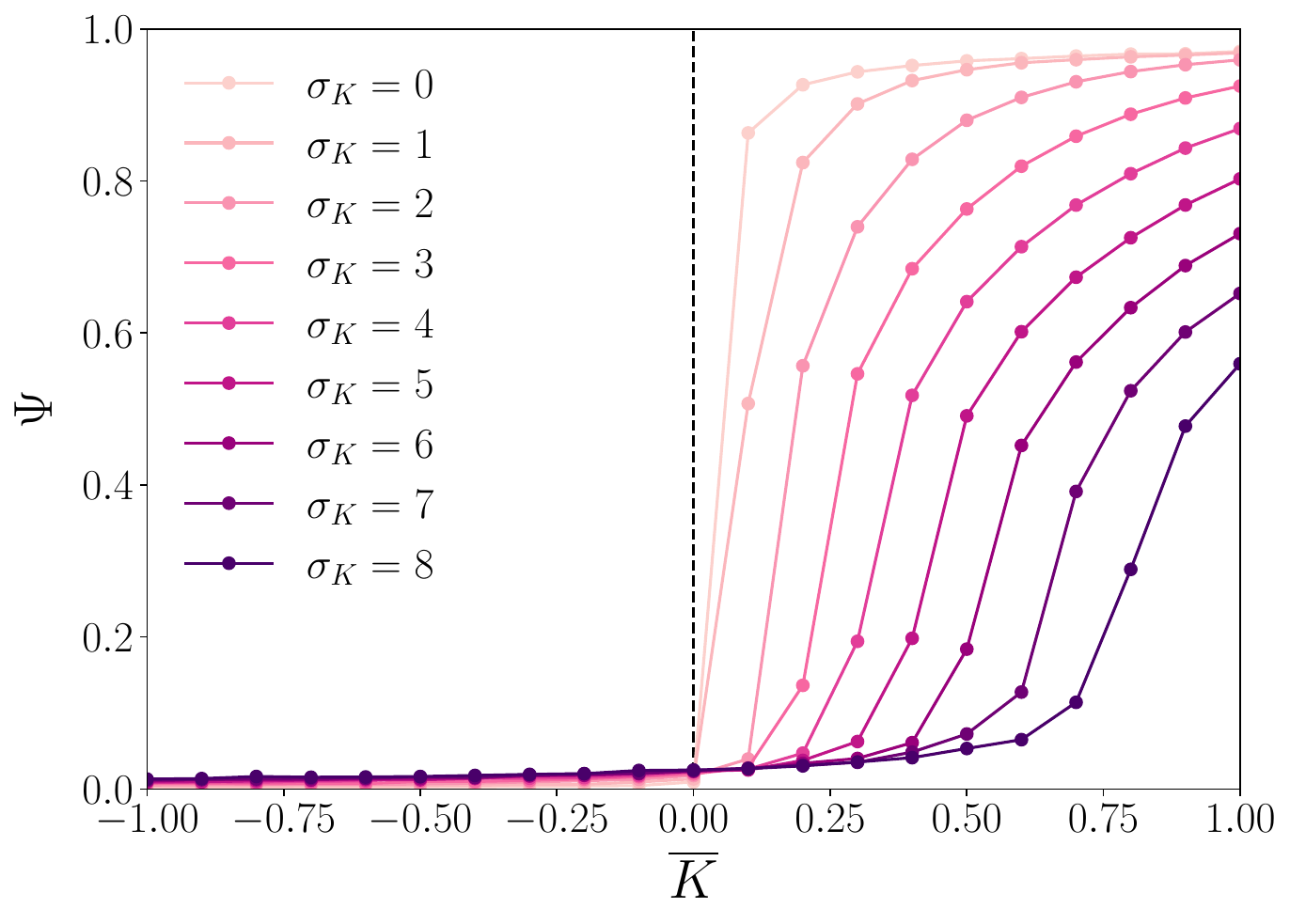}
\caption{Polar order phase transition against $\ka$ when coupling disorder is annealed for various values of $\ks$. Simulation parameters were $N=10^4$, $\rho=1$, and $\eta=0.2$.}
    \label{fig:annealed_sigma}
\end{figure*}

To support our argument, we simulated the ADXY model with annealed disorder in the couplings. The results of this were shown in the main text Fig.~3. To support this, here, we provide additional phase transition plots at a lower value of noise strength in Fig.~\ref{fig:annealed_sigma}. We clearly see that the transition point $\ka_c$ between disorder and order shifts to larger values of $\ka$ when $\ks$ is increased. This is in contrast to the quenched disorder case, where $\ka_c$ decreased when $\ks$ was increased.

\section{Mean-field calculation}

To rationalize how polar order can emerge even when the mean coupling is negative, we construct a mean-field large-deviation argument.
We have $N$ particles which interact through pairwise couplings $\kij$ i.i.d. with a Gaussian distribution such that
\begin{equation}
  P(\kij) = \frac{1}{2\pi \sigma^2} e^{-\frac{\left(\kij-\ka\right)^2}{2\ks^2}}
\end{equation}
with mean $\ka$ and standard deviation $\ks$. 
We want to know the probability of having at least one cluster made up of $N_s$ particles whose mean coupling is given by $\overline{K}_s > \epsilon$. 
This amounts to computing 
\begin{equation} \label{eq:P_binom}
  P(\overline{K}_s>\epsilon) = \frac{1}{\binom{N}{N_s}}.
\end{equation}
Within the cluster, $\overline{K}_s$ is the sum of $N_s(N_s-1)/2$ Gaussian i.i.d. random variables:
\begin{equation}
  \overline{K}_s = \frac{2}{N_s(N_s-1)} \sum_{i<j} \kij.
\end{equation}
Therefore, the distribution of $\overline{K}_s$ is still a Gaussian with mean $\ka$ and variance $\sigma_s^2 = \frac{2}{N_s(N_s-1)}\ks^2$. The probability of observing $\overline{K}_s>\epsilon$ is thus given by
\begin{equation}
  P(\overline{K}_s>\epsilon) = \frac{1}{\sqrt{2\pi \sigma_s^2}} \int_\epsilon^\infty e^{-\frac{\left(\overline{K}_s-\ka\right)^2}{2\sigma_s^2}} d\overline{K}_s = \frac{1}{2} \left[ 1- \textrm{erf}\left( \frac{\epsilon - \ka}{\sqrt{2\sigma_s^2}}\right)\right],
\end{equation}
where erf is the error function.

\begin{figure}[b!]
\centering
\includegraphics[width=0.9\textwidth]{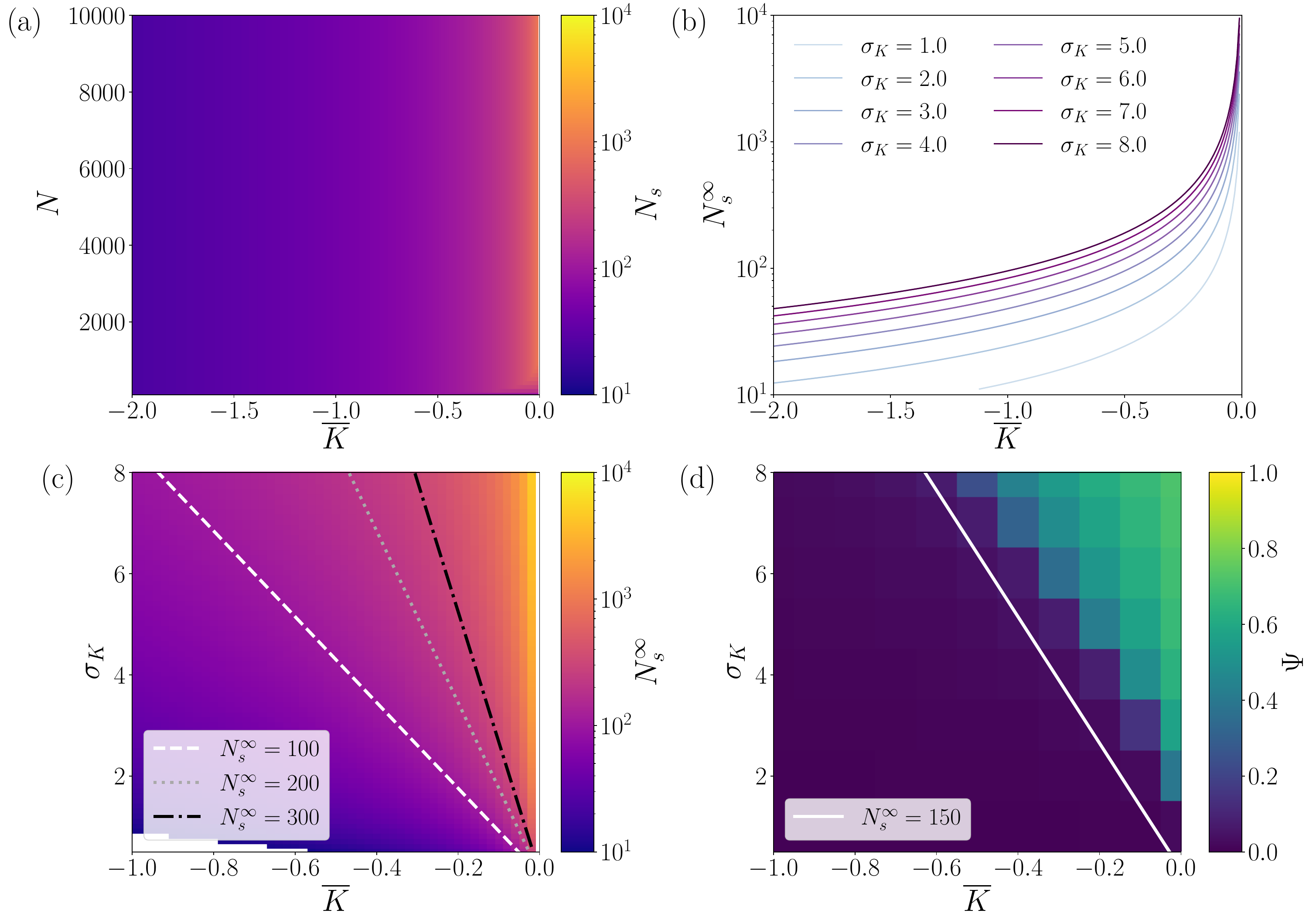}
\caption{(a) Cluster size solution $N_s$ to the self-consistency equation Eq.~\ref{eq:self_consistency} in the ($\ka$, $N$) plane at $\ks=4$, with $\eta=0.2$. The color bar for $N_s$ is logarithmic. (b) Limiting values of cluster size $N_s^\infty$ as a function of $\ka$ for different values of $\ks$.
  (c) Mean-field cluster size $N_s^\infty$ in the ($\ka$, $\ks$) plane. White regions are where the self-consistency equation does not converge to a solution $\alpha\in[0,1]$. Lines for $N_s^\infty=100$, $200$, $300$ are overlaid as dashed, dotted and dash-dotted lines, respectively. (d) Polar order parameter from simulations of the ADXY model in the ($\ka$, $\ks$) plane. The mean-field cluster size $N_s^{\infty}=150$ solution to the self-consistency equations is plotted as a white line.}
\label{fig:mf_cluster}
\end{figure}

The right hand side of Eq.~\ref{eq:P_binom} can be approximated using Stirling's approximation for $N$, assuming $N_s=\alpha N$, where $0<\alpha<1$. We then obtain the following self-consistency equation:
\begin{equation}
  \log\left\{ \frac{1}{2} \left[ 1-\textrm{erf}\left( \frac{\epsilon - \ka}{\sqrt{2\sigma_s^2}}\right) \right]\right\} = \alpha N \log \alpha + N(1-\alpha) \log(1-\alpha).
\end{equation}
Finally, we can exponentiate both sides to obtain 
\begin{equation} \label{eq:self_consistency}
  f(\alpha) \equiv \frac{1}{2} \left[ 1-\textrm{erf}\left( \frac{\epsilon - \ka}{\sqrt{2\sigma_s^2}}\right) \right] \alpha^{-\alpha N} (1-\alpha)^{-N(1-\alpha)} -1 = 0.
\end{equation}

Given $\ka$, $\sigma_K$ and $\eta$ values, we numerically solve for the root $\alpha\in(0,1)$ of $f(\alpha)=0$ for $\epsilon=\eta^2$, which corresponds to the mean-field transition threshold in the absence of coupling disorder \cite{farrell2012}, i.e., the minimum mean coupling required for a subset of particles to sustain polar order against noise.

There is no convergence for $\ka>0$ and large $N$ as a positive coupling is already favored, so it is not a rare event to find a positively aligning cluster. Therefore, we concentrate on the region $\ka<0$ where polar order is less intuitive and rare fluctuations are required to produce locally cooperative clusters. From the solution $\alpha$, we compute the cluster size $N_s=\alpha N$, and plot this in the ($\ka$, $N$) plane in Fig.~\ref{fig:mf_cluster}(a).
At a given $\ka$ and $\ks$, we find that $N_s$ remains constant for $N$ sufficiently large ($N\gtrapprox N_s$). Therefore, we choose to examine the limiting value of the cluster size $N_s^\infty$, which we plot against $\ka$ for different values of $\ks$ in Fig.~\ref{fig:mf_cluster}(b). We observe that polarized clusters are possible in the region $\ka<0$, and their size increases with $\ks$. This matches our numerical findings that quenched disorder in the couplings can promote polar order.
We also plot the mean-field cluster size $N_s^\infty$ as a heatmap in Fig.~\ref{fig:mf_cluster}(c), with the threshold lines for specific cluster sizes appearing to be linear in the ($\ka$, $\ks$) plane.

Plotting the phase diagram of polar order from simulations in the ($\ka$, $\ks$) plane in Fig.~\ref{fig:mf_cluster}(d), we find that the transition from disorder to global polar order is consistent with a critical mean-field cluster size of $N_s^\infty\approx150$.
It should be noted that the interpretation of a cluster size in the context of the mean-field calculation is very different from the spatially organized clusters within the interaction radius in simulations. In the calculation, ``clusters'' can be thought of as a collective subset of particles with mutual positive alignment capable of nucleating global alignment.

Although we have $\alpha\to0$ in the limit of large $N$, we must consider that this is a mean-field large deviation theory calculation. We assume that $\kij$ are independently distributed without any spatial organization, which is only the case at early times of the simulations. The mean-field cluster of $N_s$ particles (which remains fixed for large $N$) nucleates polar alignment, which is then amplified through activity due to spatial organization of couplings, leading to larger scale polar alignment.

\section{Density fluctuations correlations}

To quantify clustering in the system, we calculate density fluctuations correlations. Specifically, we coarse-grain the system into discrete boxes with length $\ell = 1$, and calculate local number densities $\rho(\rbf, t)$ within each box, where $\rbf$ is defined as the center of the box. We calculate the density fluctuations $\delta\rho(\rbf, t)=\rho(\rbf,t) - \langle \rho(\rbf,t)\rangle_{\rbf}$. The equal-time connected density correlation function is given by
\begin{equation}
C_\rho(r) = \frac{\langle \delta\rho(\rbf_i, t) \cdot \delta\rho(\rbf_i + \rbf, t) \rangle_{\rbf_i, t}}{\langle |\delta\rho(\rbf_i,t)|^2 \rangle_{\rbf_i, t}}.
\end{equation}
This has been normalized such that $C_\rho(0)=1$.
To speed up computation time, correlations were calculated using a Fast Fourier Transform (FFT). We plot $C_\rho$ against radial distances $r=|\rbf|$. We do not concern ourselves with measuring this in the directions perpendicular and parallel to the mean heading direction, as we are mainly concentrating on the transition between a disordered homogeneous system and small clusters appearing, which we expect to be isotropic.

\begin{figure}[h!]
  \includegraphics[width=0.5\linewidth]{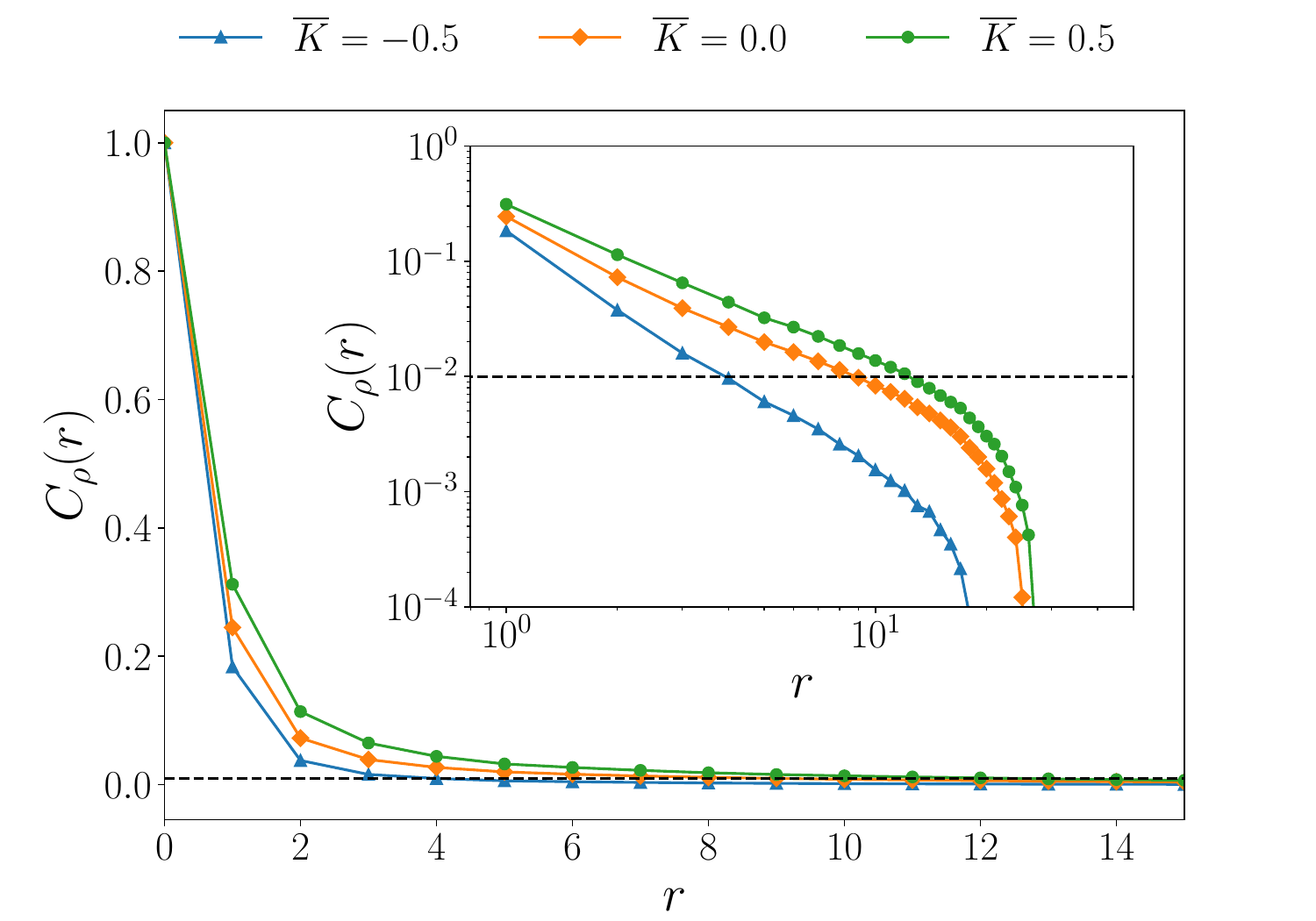}
  \caption{Density fluctuation correlation functions for $\ks=6$, $\rho=1$, and $\ka\in\{-0.5, 0.0, 0.5\}$. These are shown on a linear scale in the main figure and on a log-log scale in the inset. Correlations were averaged over time and realizations through a linear binning in $r$. A black dashed line is shown at $C_\rho(r)=0.01$ to indicate the value at which $r_c$ is defined.}
  \label{fig:corr_density}
\end{figure}

Example correlation functions are shown in Fig.~\ref{fig:corr_density} at fixed $\ks>0$ and $\rho$ for increasing $\ka$, across the order-disorder transition.
When plotted on a log-log scale, we see that for intermediate values of $r$, these correlation functions follow a power law scaling. We further observe an exponential cutoff for large values of $r$ due to finite size effects. As described in the main text, we define $r_c$ to be the radial distance such that density fluctuations correlations have decayed enough, i.e. $C_\rho(r_c)=0.01$, and use this to estimate the typical cluster size in the system. This value was chosen as it signifies that the correlations have sufficiently decayed, but is also measured in the algebraic scaling regime so that measurements are not skewed by the exponential cut-off.

As $\ka$ increases, the value of $r_c$ increases, indicating the size of clusters is increasing. Despite the system being disordered at $\ka =-0.5$ ($\Psi\approx0$), the system still shows clustering with $r_c\approx4$, defining a \textit{disordered} but \textit{clustered} phase.

\section{Hyperuniformity in the ADXY model}

Motivated by our findings of regions of larger density fluctuations and clustering within the disordered phase of the model, we study further the statistics of density fluctuations. In line with literature on hyperuniformity \cite{torquato2018}, we measure the variance of the local density $\sigma^2(\rho_\ell)$, where $\rho_\ell$ is the density of particles found in the region of volume $V = \ell^d$. The average is taken over many different volumes of the same system as well as different configurations. 

We expect the variance of the local density to scale with the size of the probing window such that
\begin{equation}
\sigma^2(\rho_\ell) \sim \ell^{-\lambda}
\end{equation}
For Poisson point processes, we expect that the scaling exponent $\lambda = d$ (here, $d=2$). If $\lambda>d$, then the density fluctuations are decaying faster than for a uniform system, such a system is called hyperuniform. Finally, if $\lambda<d$, then one observes enhanced density fluctuations, the system is structured in space, and we can call our system clustered.

\begin{figure}[h!]
	\includegraphics[width=\linewidth]{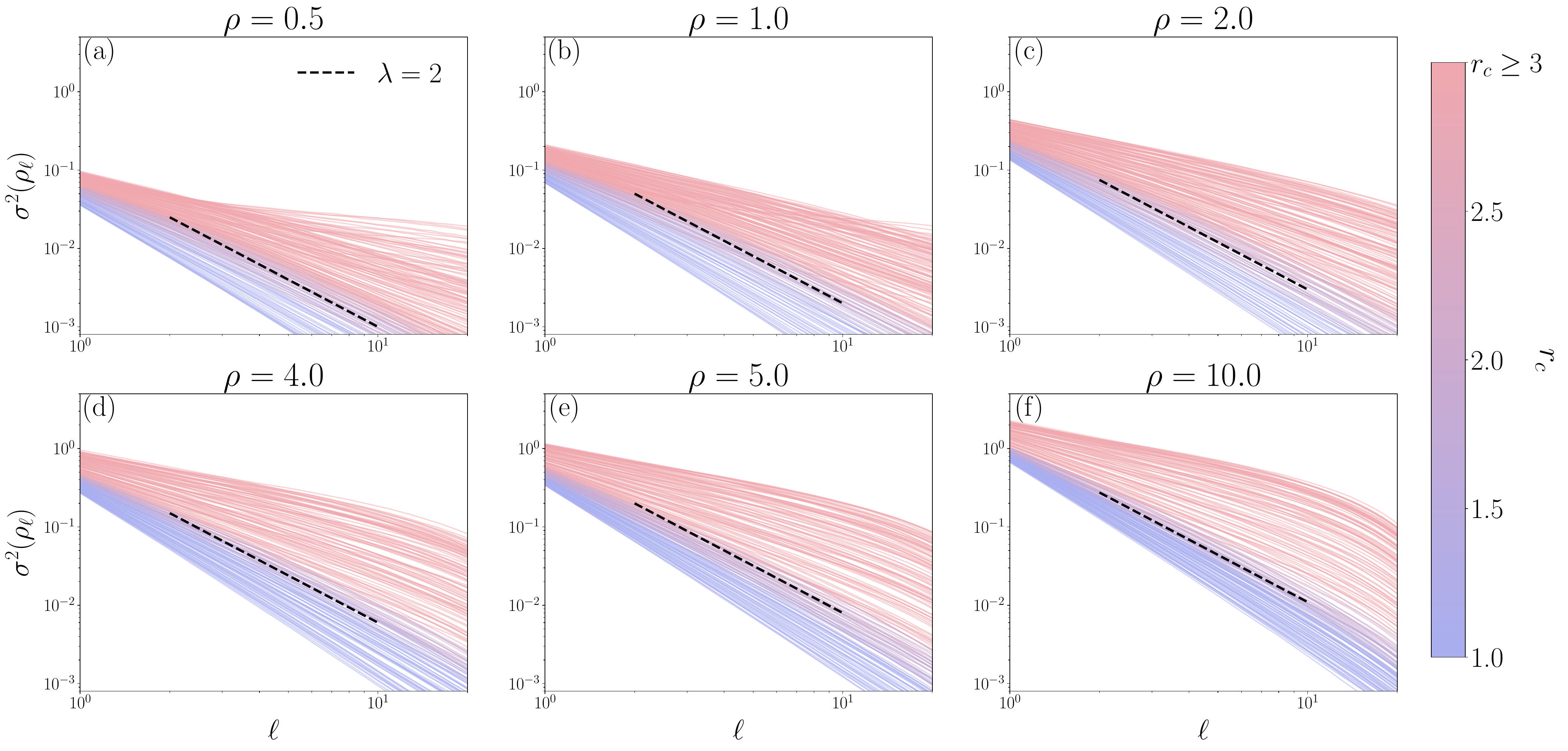}
	\caption{Density variance $\sigma^2(\rho_\ell)$ vs $\ell$ for each of $\rho\in\{0.5, 1, 2, 4, 5, 10\}$ in (a)-(f). In each subplot, the lines are plotted for $\ka\in\{-1.0, -0.9, \dots, 1.0\}$ and $\sigma_K\in\{0,1,2,\dots,8\}$. Each curve is here colored according to the value of $r_c$ as measured from the density correlation functions. A dashed black line is shown for exponent $\lambda=2$. Other simulation parameters were $N=10^4$, $\eta=0.4$, and $10^4$ sample observation windows of the final simulation snapshot in steady state were taken for 20 realizations for each value of $\ell$ to obtain the variance.}
	\label{fig:density_var_vs_ell}
\end{figure}

\begin{figure}[t]
	\includegraphics[width=0.6\linewidth]{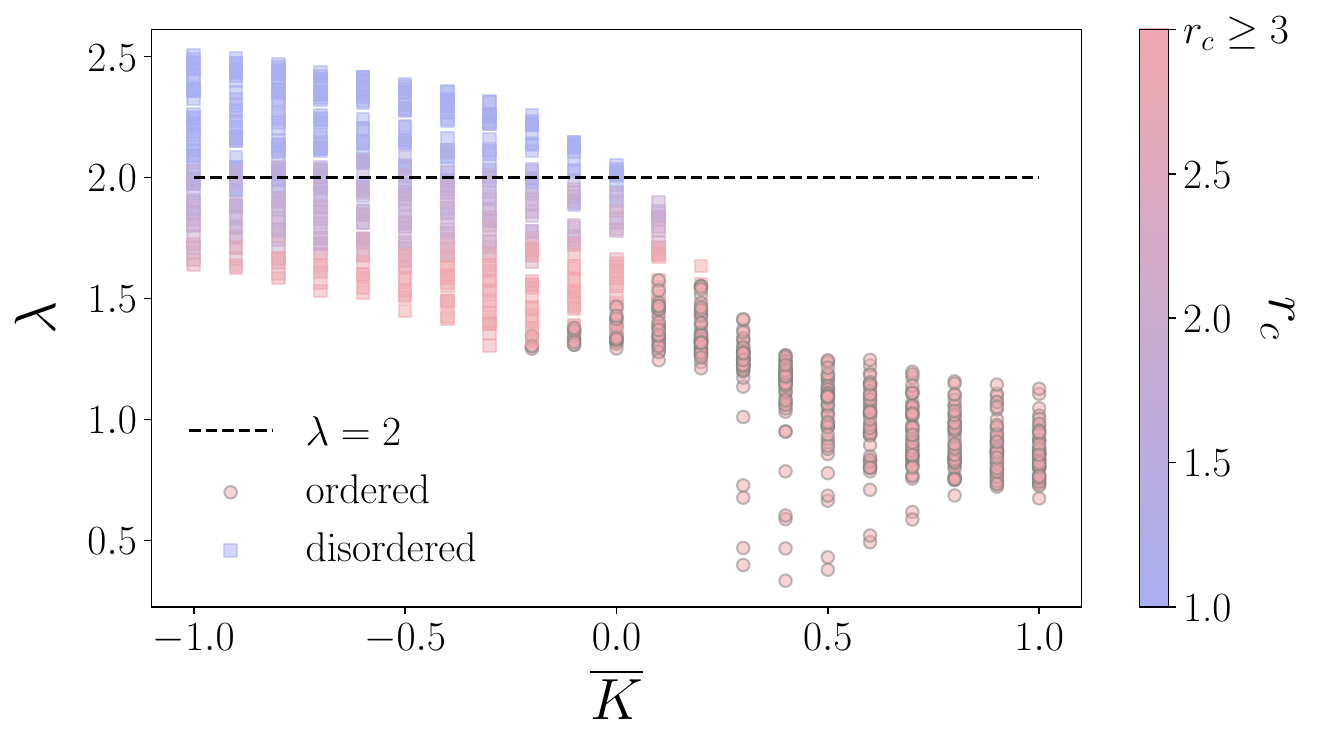}
	\caption{Power law exponent fitted from density variance $\sigma^2(\rho_\ell)\sim\ell^{-\lambda}$ for $2\leq \ell \leq 10$ as a function of $\ka$. All points are plotted with $\rho\in\{0.5, 1, 2, 4, 5, 10\}$ and $\sigma_K\in\{0,1,2,\dots,8\}$. Each point is here colored according to its value of $r_c$ as measured from the density correlation functions. Squares are obtained in the disordered phase ($\Psi<0.1$) and circles are configurations in the ordered phase ($\Psi >0.1$).  A dashed black line shows the exponent $\lambda=2$ for uniform systems. Other simulation parameters were $N=10^4$, and $\eta=0.4$.}
	\label{fig:density_var_exp_l2-10}
\end{figure}

As can be seen in Fig.\,\ref{fig:density_var_vs_ell}, the configurations obtained in the disordered phase display density fluctuations characterized by an exponent $\lambda$ varying from $\lambda>2$ to $\lambda<2$ as the typical cluster size as measured from the density correlation functions increases. We confirm the existence of a transition from {\it hyperuniform} to {\it clustered} configurations in the disordered phase by plotting in Fig.\,\ref{fig:density_var_exp_l2-10} the values of $\lambda$ measured via a power-law fit. 

\section{Distribution of pairwise contact times}
We define the pairwise contact time to be the total number of snapshots in which particles $i$ and $j$ are within distance $r_{\textrm{max}}$ of each other, over a certain period of time. In this way, we can define the percentage of time in contact as 
\begin{equation}
\tau_{ij} = \sum_{t=1}^{N_T}H(r_{\textrm{max}} - |\rbf_i(t)-\rbf_j(t)|)/N_T
\end{equation}
where $N_T$ is the total number of snapshots, and $H(r)$ is the Heaviside step function. By construction, our pairwise contact time lies in the range $\tau\in[0,1]$.

\begin{figure}[h!]
\includegraphics[width=0.95\linewidth]{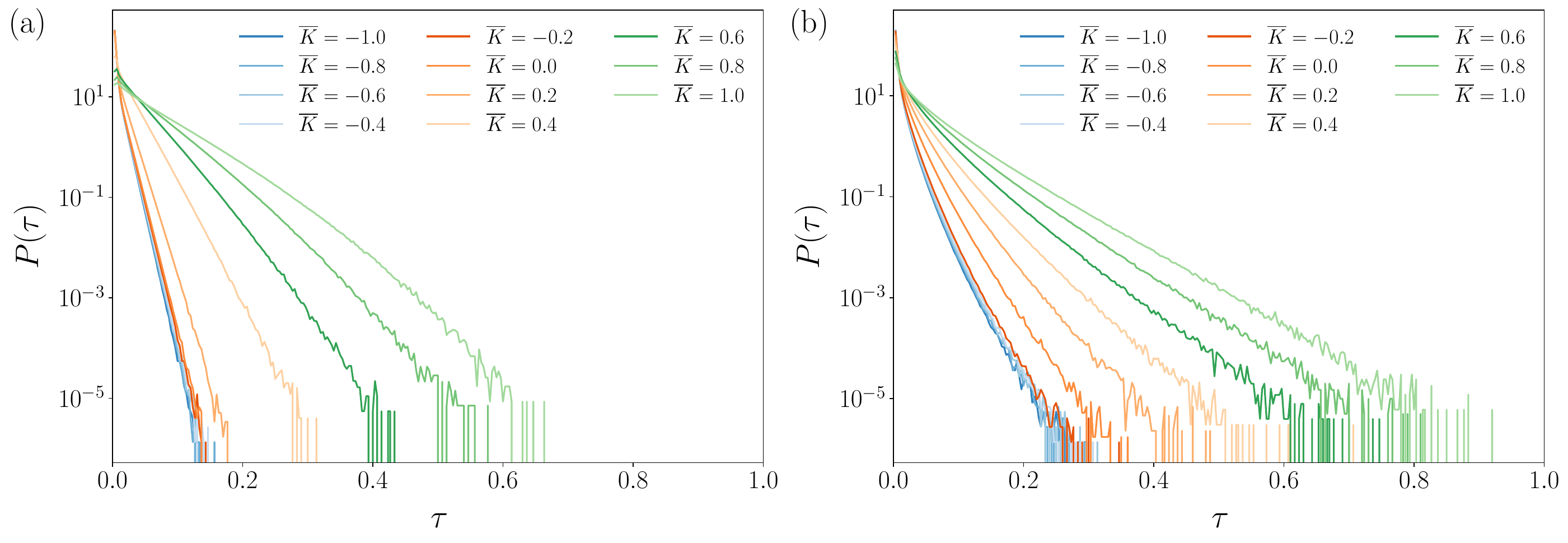}
\caption{Pairwise contact time distributions for (a) $\sigma_K=0, \ \rho=4$ and (b) $\sigma_K=6, \ \rho=4$.}
\label{fig:contact_dist}
\end{figure}

\begin{figure}[h!]
  \includegraphics[width=0.5\linewidth]{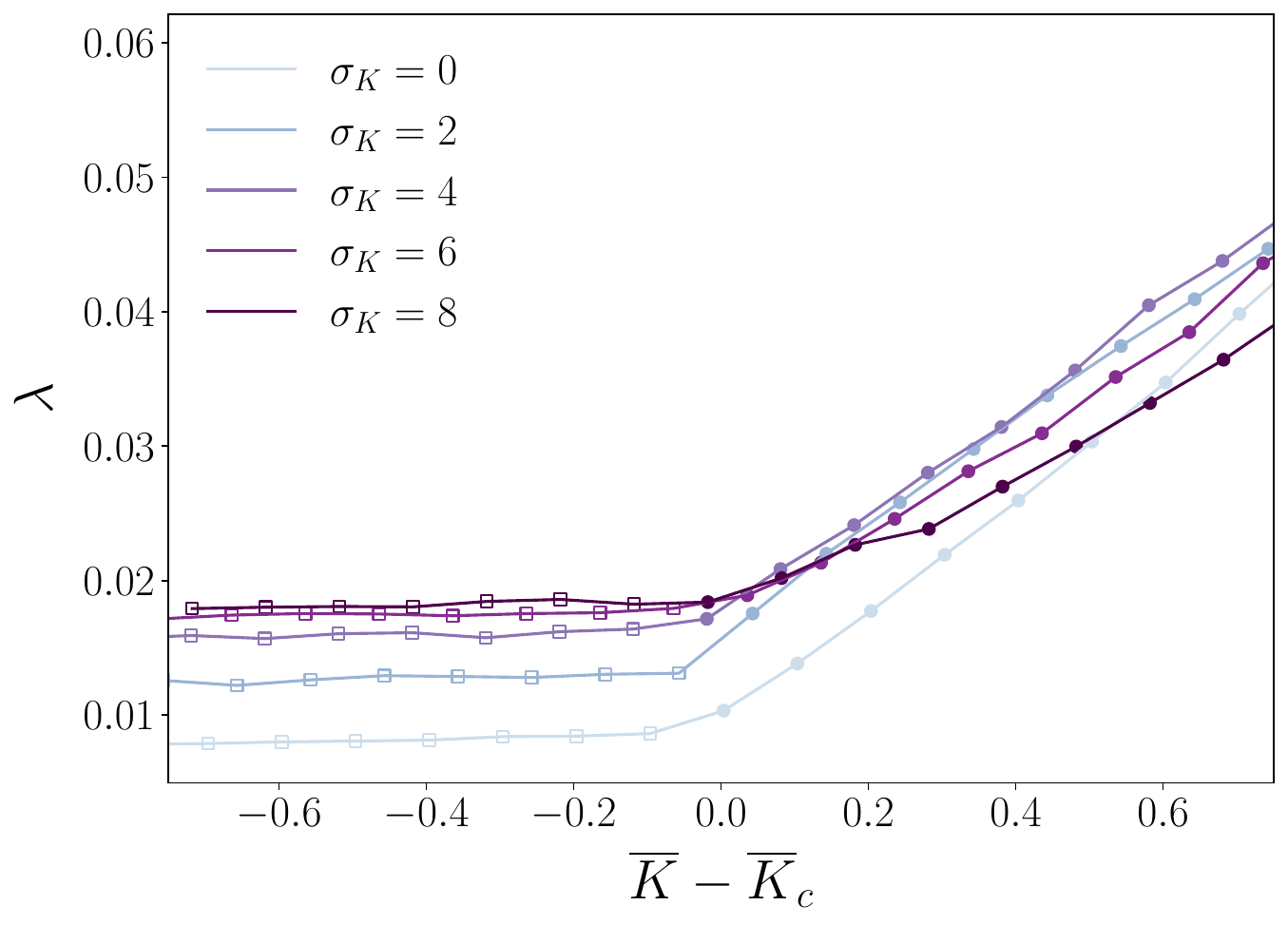}
  \caption{Exponent measured from the probability distributions of pairwise contact times $P(\tau)$ as a function of the distance to the critical average coupling strength $\overline{K}_c$. To obtain $\lambda$, exponential functions of the form $\sim e^{-\tau/\lambda}$ were fitted between the 20th and 80th percentile of unique nonzero $\tau$ values measured.
  Empty squares denote systems with no polar order ($\Psi<0.1$) and filled circles denote systems with polar order ($\Psi\geq0.1$), in line with the phase diagram in main text Fig.~4. Simulations were performed with $N=10^4, \ \rho=4, \ \eta=0.4, \ \ks=6$, and $r_{\textrm{max}}=1$.}
  \label{fig:contact_exponent}
\end{figure}

We find that the pairwise contact times as defined above are distributed exponentially (see Fig.~\ref{fig:contact_dist}). By fitting the distributions to an exponential distribution $e^{-\tau/\lambda}$, we can extract from the data typical pairwise contact times. We measure over intermediate values of $\tau$ to avoid any skew from very short random scattering events as well as outliers with large contact times. The exponents are plotted for $r_{\textrm{max}}=1$ in Fig.~\ref{fig:contact_exponent}. Here, we denote $\ka_c$ the value of the mean coupling strength at which the system transitions from disordered to ordered. We find that pairwise contact time distributions remain unchanged throughout the disordered phase ($\overline{K}-\overline{K}_c<0$); in the ordered phase  ($\overline{K}-\overline{K}_c>0$), the typical pairwise contact time increases as the mean coupling strength increases for all values of $\ks$.

\begin{figure}[h!]
	\includegraphics[width=0.5\linewidth]{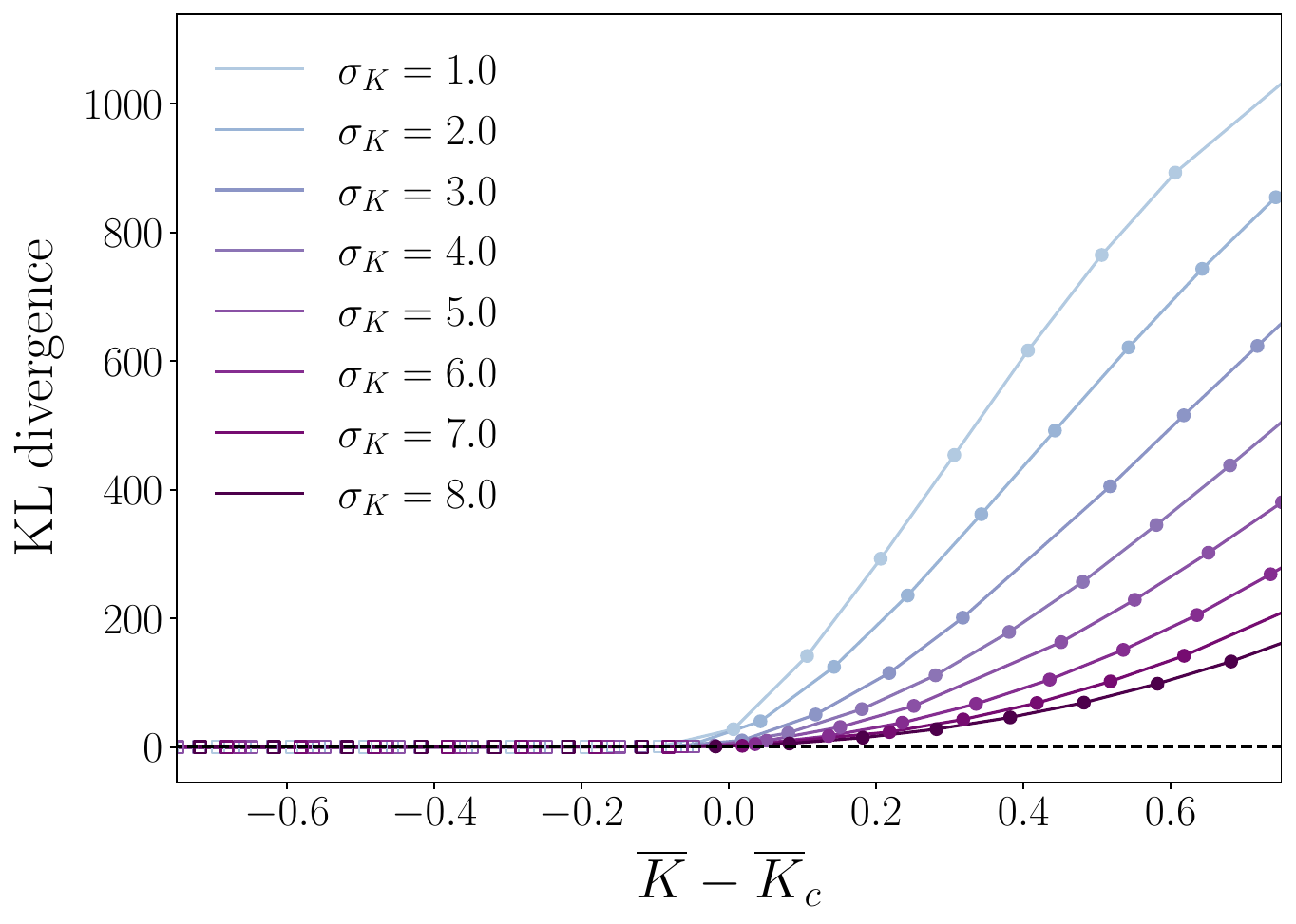}
	\caption{KL divergence for percentage contact time probability distributions $P(\tau)$ compared against the baseline of $\ka=-1.0$ and the same value of $\ka$. Empty squares denote systems with no polar order ($\Psi<0.1$) and filled circles denote systems with polar order ($\Psi\geq0.1$), in line with the phase diagram in main text Fig.~4. Simulations were performed with $N=10^4, \ \rho=4, \ \eta=0.4$, and $r_{\textrm{max}}=1$.}
	\label{fig:contact_kl}
\end{figure}

As a further measure to compare the probability density functions of contact times, we measure the Kullback-Leibler (KL) divergence for $P_{\ka,\ks}(\tau)$ using as reference the distribution observed for the system furthest into the disordered phase at a given $\ks$ ($\ka=-1$ here). Here, the KL divergence is thus defined as
\begin{equation}
  D_{KL}(\ka, \ks) = \sum_{\tau, P(\tau)\neq0} P_{\ka,\ks}(\tau) \log\left(\frac{P_{\ka,\ks}(\tau)}{P_{-1,\ks}(\tau)}\right) 
\end{equation}
where we stipulate that the sum must only be measured over percentage contact times $\tau$ such that $P_{-1,\ks}(\tau)$ and $P_{\ka,\ks}(\tau)$ are both nonzero.
The divergence is zero if the distributions are identical over this range of $\tau$, and positive nonzero otherwise. We clearly see in Fig.~\ref{fig:contact_kl} that the KL divergence remains very close to zero for all values of mean coupling strength such that $\overline{K}-\overline{K}_c<0$. The KL divergence then increases monotonically with mean coupling strength in the ordered phase. 


\section{Banding persists in the ADXY model}

As noted in the main text, we still observe a coexistence phase in the ADXY model in which the system displays ordered bands. These are, however, subject to large finite-size effects. To confirm their existence, we had to simulate large systems in a slab geometry when $\ks>0$.

\begin{figure}[!h]
  \includegraphics[width=\linewidth]{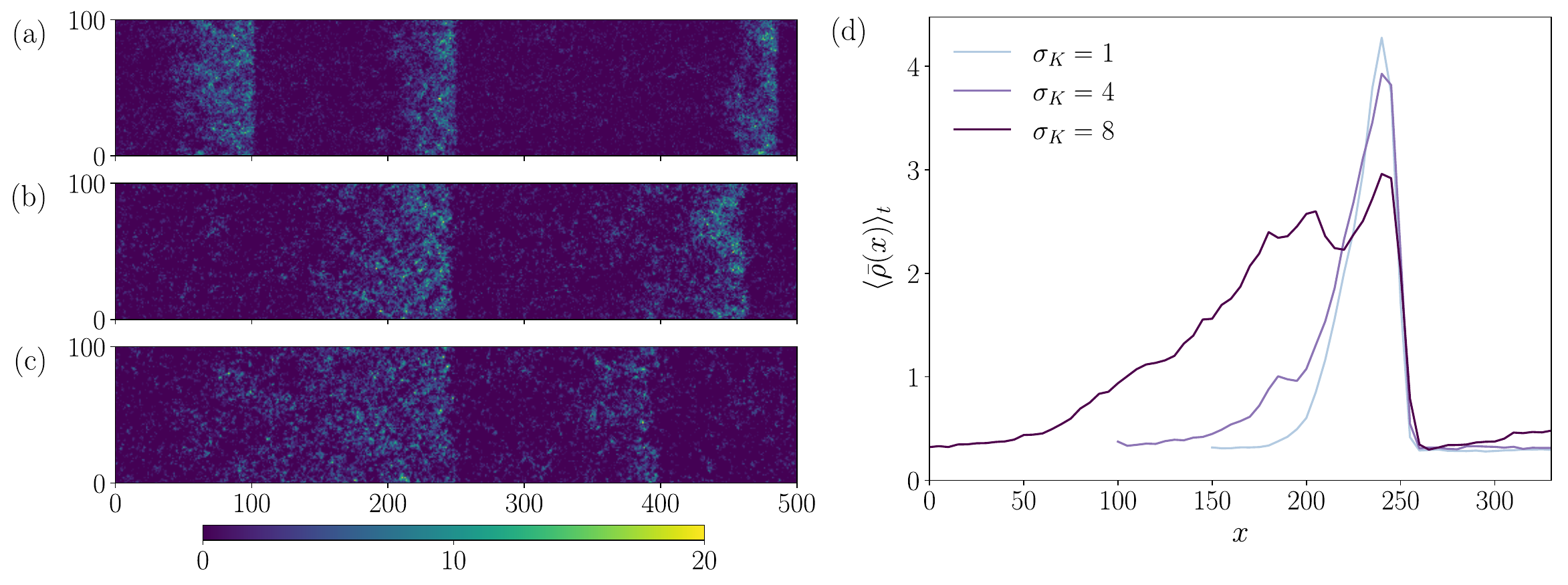}
  \caption{Banding for increasing $\ks$. (a)-(c) Snapshots of the coarse-grained densities in the coexistence phase for $\ks=1$, $4$ and $8$, respectively. (d) Density profiles perpendicular to the direction of travel, averaged over time at intervals of $t=10$ from $t=5000$ to $6000$.
  Simulations were performed with $N=5\times10^4$, $L_x=500$, $L_y=100$, $\rho=1$, $\eta=0.7$, $\ka=1$.}
  \label{fig:bands_n0.7}
\end{figure}

In Fig.~\ref{fig:bands_n0.7}, we analyze the effect of increasing $\ks$ on the bands, at fixed $\ka$ and $\rho$. We have used a large noise value of $\eta=0.7$, as bands are observed for all values of $\ks$ studied here; in this case, we find clear and stable bands and are able to average over their transverse density profiles, shown in Fig.~\ref{fig:bands_n0.7}(d). As $\ks$ increases, the band wavelength increases, and the density inside the band decreases. The density of the bands needs to decrease in order to sufficiently spread out the highly aligning clusters. Otherwise, this would cause frustration and break apart the bands into a disordered phase.

\begin{figure}[!h]
  \includegraphics[width=\linewidth]{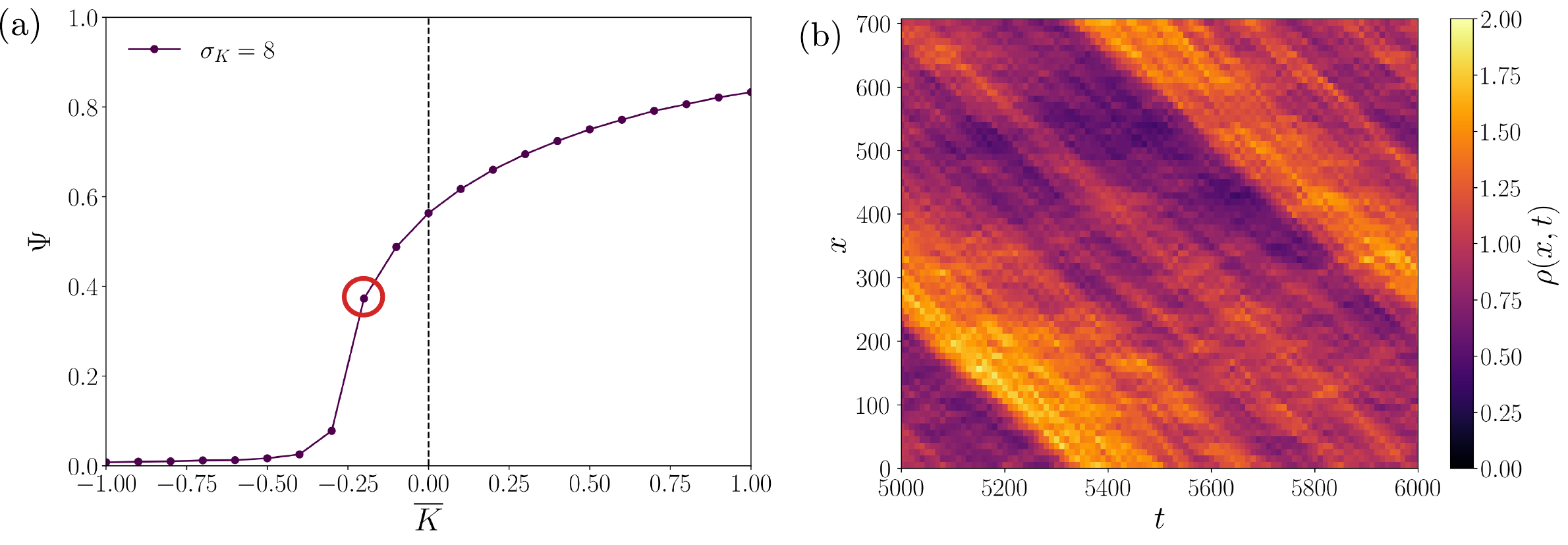}
  \caption{Banding for $\ka<0$. (a) Polar order parameter $\Psi$ against $\ka$ for $N= 10^4, \ \rho=1, \ \eta=0.2$, and $\ks=8$. The point at $\ka=-0.2$ is circled in red to highlight where we look for banding. (b) Kymograph of the local density in slices transverse to the direction of travel against time. This shows a clear, linear periodic band moving in time. Simulations for (b) were performed with $N=10^5$ and $L_x/L_y=5$.}
  \label{fig:band_kymograph_n0.2}
\end{figure}

At $\eta=0.7$, the transition to polar order occurs at $\ka>0$. We therefore wish to verify that bands are still observed in polar ordered regions close to the transition when $\ka_c<0$, which we can achieve by decreasing $\eta$ and making $\ks$ large enough. We choose $\eta=0.2$ to ensure the region $\ka<0$ is sufficiently deep in the banding phase to see clear bands.
However, as noise decreases, the difference in density within and outside the bands becomes even smaller. At the same time, we saw that a higher $\ks$ causes the bands to spread out and wavelength increases. We therefore plot kymographs of the transverse density against time to provide a clearer picture for the moving bands when $\ks$ is large and $\eta$ is small.
With large enough systems, we can see clear signatures of banding in the kymograph when $\ka<0$, see Fig.~\ref{fig:band_kymograph_n0.2}.

\begin{figure}[b!]
  \includegraphics[width=\linewidth]{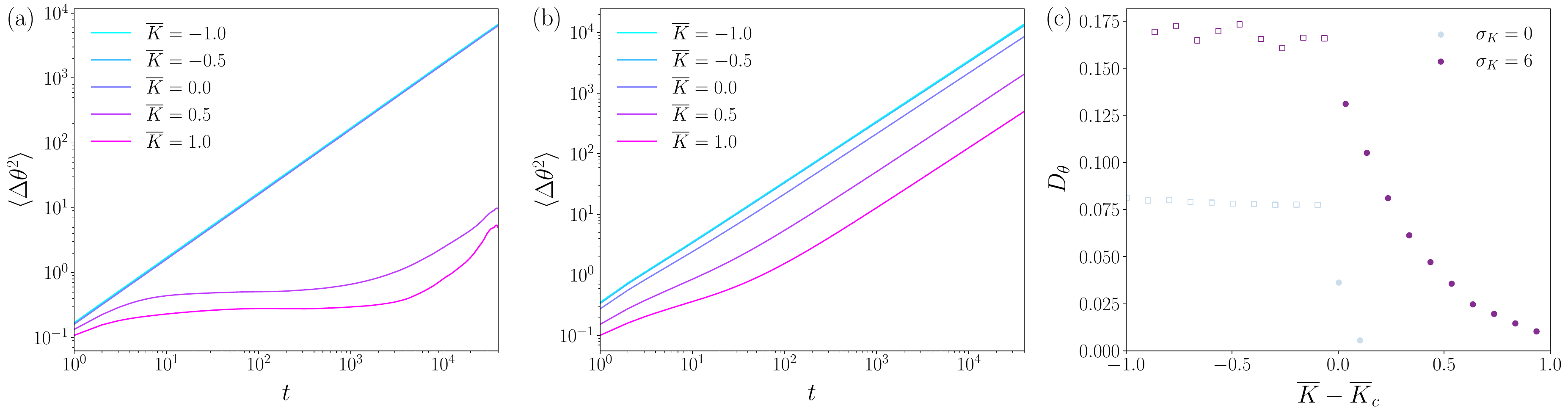}
  \caption{(a) Mean angular MSD for various values of $\ka$ for fixed $\ks=0$. (b) Mean angular MSD for various values of $\ka$ for fixed $\ks=6$. (c) Effective diffusion coefficient by measuring $D_\theta$ from a linear fit over the interval $t\in[10^3, 10^4]$ for various values of $\ka - \ka_c$ and $\ks=0,\ 6$. Empty squares denote systems with no polar order ($\Psi<0.1$) and filled circles denote systems with polar order ($\Psi\geq0.1$), in line with the phase diagram in main text Fig.~4. Simulations were performed with $N=10^4, \ \rho=4, \ \eta=0.4, \ \ks=6$, and were run to $t=4\times 10^4$. }
  \label{fig:angular_msd}
\end{figure}

\section{Angular MSD}
In this section, we measure a time-averaged angular mean squared displacement (MSD) for each particle, before averaging over particles, realizations and time for each realization
\begin{equation}
  \left\langle\Delta \theta^2 \right\rangle (t) = \left\langle \frac{1}{N}\sum_{i=1}^N \left(\theta_i(t+\tau) - \theta_i(\tau)\right)^2 \right\rangle_\tau
\end{equation}

For sufficiently large $t$, we expect to observe rotational diffusion consistent with the following scaling $\langle\Delta \theta^2\rangle=2D_\theta t$, where $D_\theta$ is the effective angular diffusion coefficient. We plot both the angular MSD over time and the long-time effective diffusion coefficients in Fig.~\ref{fig:angular_msd}, comparing $\ks=0$ and $\ks=6$. By varying $\ka$ across the polar order-disorder transition, we see that $D_\theta$ is roughly constant within the disordered phase and begins to decrease at the transition point as we enter the polar ordered phase for both $\ks=0$ and $\ks=6$. We also observe that diffusion coefficients are uniformly smaller for $\ks = 0$ across the whole range of $\ka$ tested. We thus conclude that flocks are more fragile (particles are overall less persistent in direction) when quenched disorder is added in the interaction couplings.

\section{Self-averaging}
Previous work has shown that different kinds of disorder in a flying XY model can break the self-averaging property of a system \cite{duan2021}. This is an important quantity to check to ensure all averages taken over time (in steady state) and realizations are valid. To test for the presence of self-averaging in the ADXY model on approaching the transition point, for different system sizes, we measure the relative variance of the polar order parameter
\begin{equation}
  R_\Psi = \frac{\langle \Psi^2 \rangle - \langle \Psi \rangle^2}{\langle \Psi \rangle^2},
\end{equation}
where the averages in angular brackets are taken over time and realizations. Away from the transition point, in the ordered phase, this quantity should scale inversely with system size $R_\Psi\propto N^{-1}$. If this scaling persists up to the transition point then the system is strongly self-averaging. If $R_\Psi\propto N^{-z}$ with $0<z<1$, then the system is weakly self-averaging. If $R_\Psi$ decays to a constant as $N\to\infty$ close to the transition point, then the system is not self-averaging.

\begin{figure}[b!]
  \includegraphics[width=\linewidth]{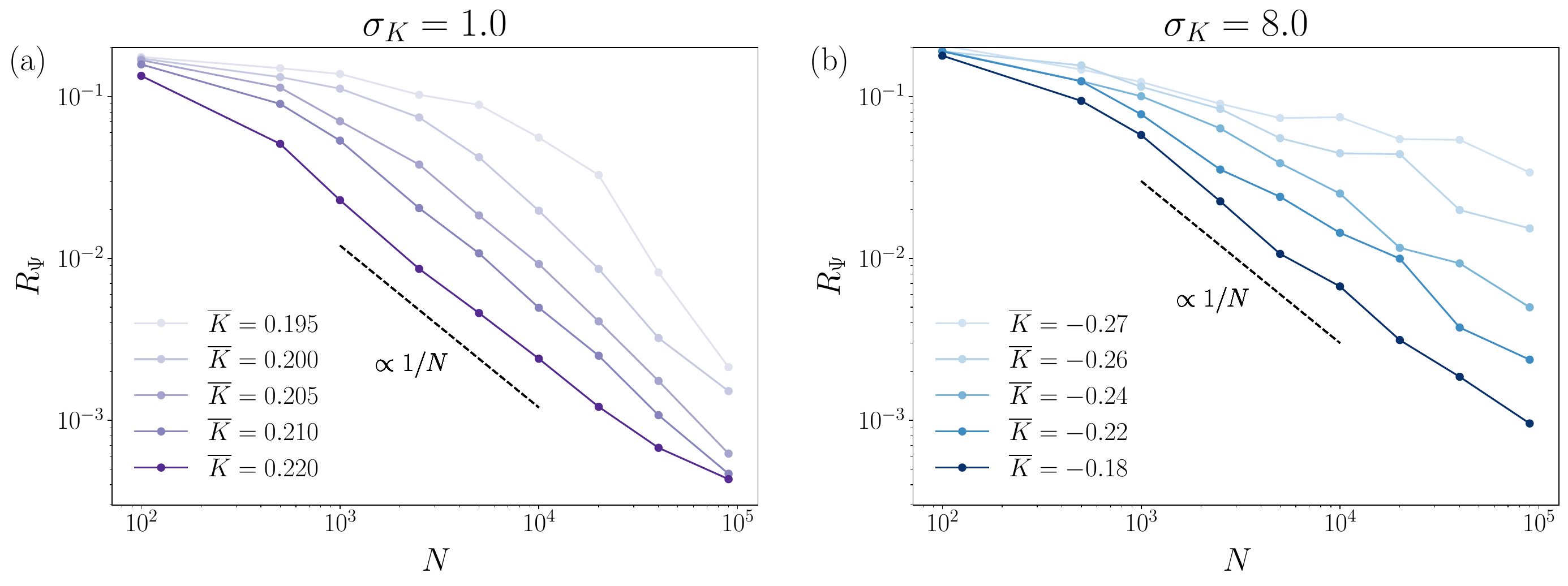}
  \caption{Relative variance against system size, for different values of $\ka$, approaching the transition point for (a) $\ks=1$ ($\ka_c=0.193(7)$) and (b) $\ks=8$ ($\ka_c=-0.282(0)$). Simulations were performed with $\rho=1$, $\eta=0.4$, and averaged over 100 realizations.}
  \label{fig:self_average}
\end{figure}

We plot the relative variance for the ADXY model with small and large values of $\ks$ in Fig.~\ref{fig:self_average}, where we simulated system sizes over three decades in $N$ up to a maximum system size of $N=10^5$. We start in the polar ordered state, and slowly decrease $\ka$ towards the transition point $\ka_c$.

We indeed find that $R_\Psi$ scales as $N^{-1}$ at $\ka\gg \ka_c$. As the value of $\ka$ approaches $\ka_c$, we find that $R_\Psi$ still decays for large $N$ for both small and large $\ks$, at least for these system sizes. We conclude that the system is self-averaging. The rate of decay decreases on approaching $\ka_c$ for $\ks=8$; however, given the strong finite size effects of the model for large $\ks$, we would need much larger system sizes to confirm whether the ADXY model is weakly or strongly self-averaging.

\section{Frustration}

We now quantify frustration in the system. First, we note that the mean interaction strength $\langle K_{ij} \rangle_{\sigma_I}$---where the average is taken over all particle pairs within distance $\sigma_I$---increases with $\ks$ [Fig.~2(d) in the main text]. However, this does not imply that all antiferromagnetic couplings ($\kij<0$) become irrelevant. As $\ks$ increases, a significant proportion of interactions remain frustrated, in the sense that they contribute unfavorably to the energy.

For each interacting pair we define the contribution to the Hamiltonian,  
\begin{equation}
E_{ij} = -\kij \cos(\theta_j-\theta_i).
\end{equation}  
A pair is considered \emph{frustrated} if $E_{ij}>0$, i.e. if the interaction increases the energy. We therefore introduce a measure of \emph{relative frustration} as the ratio of unfavorable to total interactions, averaged over particles, time (after steady state), and realizations:  
\begin{equation}
f = \frac{\sum_{i, j\in\mathcal{N}_i, t}\mathds{1}_{E_{ij}(t)>0}}{\sum_{i,t} n_i(t)},
\end{equation}
where $\mathds{1}$ is the indicator function and $n_i(t)$ is the number of neighbors of particle $i$ at time $t$. Thus, $f=0$ corresponds to a fully frustration-free system, while $f>0$ indicates the presence of unfavorable interactions.  A value of frustration, $f=0$, would indicate that there is no frustration in the system, as all interactions are favorable. 

\begin{figure*}[b!]
\centering
\includegraphics[width=0.5\linewidth]{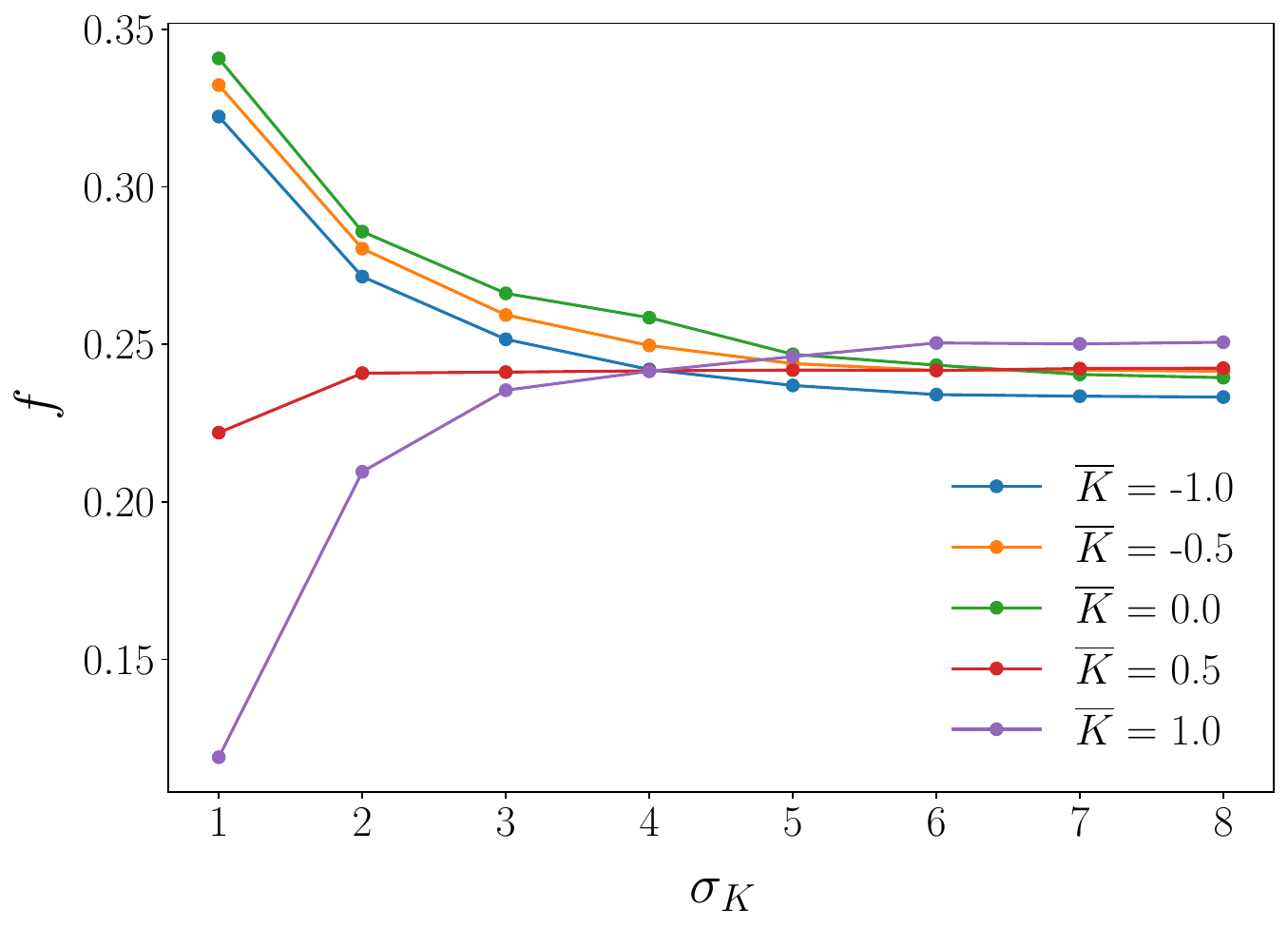}
\caption{
Relative frustration $f$ against $\ks$ for various $\ka$.}
\label{fig:frustration}
\end{figure*}

Figure~\ref{fig:frustration} shows that for $\ka=1.0$ and $\ka=0.5$, frustration increases monotonically with $\ks$. By contrast, for $\ka=0.0, -0.5, -1.0$, frustration decreases as $\ks$ grows. The crossover coincides with the value of $\ka$ where increasing $\ks$ changes from promoting to suppressing global polar order (around $\ka \approx 0.3$ for the parameters used here; see Fig.~1 in the main text).

To gain intuition, consider the system at random initial conditions, where angles are uniformly distributed and hence independent of the couplings. In this case,  
\begin{equation}
\mathbb{P}(E_{ij}>0) = \mathbb{P}(\kij\cos(\theta_j-\theta_i)<0).
\end{equation}
Independence implies  
\begin{equation}
\mathbb{P}(E_{ij}>0) = \mathbb{P}(\kij>0)\,\mathbb{P}(\cos(\theta_j-\theta_i)<0) + \mathbb{P}(\kij<0)\,\mathbb{P}(\cos(\theta_j-\theta_i)>0).
\end{equation}
The difference in angles $\theta_j-\theta_i$ is clearly uniform on $[0,2\pi]$. Although the distribution of $\cos{(\theta_j-\theta_i)}$ will no longer be uniform, it will crucially be symmetric about zero. Indeed, since cosine is an even function, the PDF of the cosine of a random variable will also be an even function. We can therefore write
\begin{equation}
\mathbb{P}(\cos(\theta_j-\theta_i)>0) = \mathbb{P}(\cos(\theta_j-\theta_i)<0) = \tfrac{1}{2}.
\end{equation}
Therefore, for any Gaussian distribution $\kij \sim \mathcal{N}(\ka, \ks^2)$ with $\ks>0$, we obtain  
\begin{equation}
\mathbb{P}(E_{ij}>0) = \tfrac{1}{2}.
\end{equation}
This sets an upper bound for the frustration, consistent with our simulations, which always yield $f<1/2$. Thus, the system dynamically alleviates some frustration, but not all of it, especially as $\ks\to\infty$.

Interestingly, as $\ks$ increases, all datasets converge to $f \approx 0.25$, independent of $\ka$. This shows that even at large disorder, the system retains a finite fraction of unfavorable interactions, i.e. complete frustration relief is impossible. Unlike a conventional spin glass, where disorder is quenched and frustration is unavoidable, here particles are mobile and able to reorganize to reduce frustration. Nevertheless, the system settles into local energy minima where a residual frustration of about $25\%$ persists. Whether this alleviation suffices to sustain global polar order depends sensitively on the parameters, in particular on the value of $\ka$.

\bibliography{asg-refs}